\documentclass[prb,twocolumn,superscriptaddress,floatfix,letterpaper,longbibliography]{revtex4-1}
\pdfoutput=1

\usepackage[all]{xy}

\usepackage{euscript}
\newcommand\Rep{\EuScript{R}\mathrm{ep}}
\newcommand\sRep{\mathrm{s}\EuScript{R}\mathrm{ep}}
\newcommand{\Ve}{\EuScript{V}\mathrm{ec}}
\newcommand{\sVe}{\mathrm{s}\EuScript{V}\mathrm{ec}}

\renewcommand{\myfrm}[1]{ \medskip\\ \noindent \textit{#1}  \medskip\\}

\newcommand\sA{\EuScript{A}}
\newcommand\sC{\EuScript{C}}
\newcommand\sD{\EuScript{D}}
\newcommand\sE{\EuScript{E}}
\newcommand\sM{\EuScript{M}}

\DeclareMathOperator{\Hom}{Hom}
\DeclareMathOperator{\id}{id}

\DeclareMathOperator{\Aut}{Aut}

\newcommand{\Sq}{\mathrm{Sq}}

\newcommand\hcup[1]{\underset{{\scriptscriptstyle #1}}{\smile}}

\begin{document}

\title{A classification of 3+1D bosonic topological orders (II):\\
the case when some pointlike excitations are fermions
}

\author{Tian Lan} 

\affiliation{Institute for Quantum Computing,
  University of Waterloo, Waterloo, Ontario N2L 3G1, Canada}

\author{Xiao-Gang Wen}
\affiliation{Department of Physics, Massachusetts Institute of
Technology, Cambridge, Massachusetts 02139, USA}

\begin{abstract} 
In this paper, we classify EF topological orders for 3+1D bosonic systems where
some emergent pointlike excitations are fermions.  (1) We argue that all 3+1D
bosonic topological orders have gappable boundary.  (2) All the pointlike
excitations in EF topological orders are described by the
representations of $G_f=Z_2^f\leftthreetimes_{e_2} G_b$ -- a $Z_2^f$ central
extension of a finite group $G_b$ characterized by $e_2\in H^2(G_b,Z_2)$.  (3)
We find that the EF topological orders are classified by 2+1D anomalous
topological orders $\EuScript{A}_b^3$ on their unique canonical boundary.  Here
$\EuScript{A}_b^3$ is a unitary fusion 2-category with simple objects labeled
by $\hat G_b=Z_2^m\leftthreetimes G_b$.  $\EuScript{A}_b^3$ also has one
invertible fermionic 1-morphism for each object as well as
quantum-dimension-$\sqrt 2$ 1-morphisms that connect two objects $g$ and $gm$,
where $g\in \hat G_b$ and $m$ is the generator of $Z_2^m$.  (4) When $\hat G_b$
is the trivial $Z_2^m$ extension, the EF topological orders are called EF1
topological orders, which is classified by simple data $(G_b,e_2,n_3,\nu_4)$,
where $n_3\in H^3(G_b,Z_2)$, and $\nu_4$ is a 4-cochain
in $C^4(G_b,U(1))$ satisfying $\text{d} \nu_4=(-)^{n_3\smile n_3+e_2\smile n_3}$.  
(5) When $\hat G_b$ is a non-trivial $Z_2^m$ extension, the EF
topological orders are called EF2 topological orders, where some intersections
of three stringlike excitations must carry Majorana zero modes.  
(6) Every EF2 topological order
with $G_f=Z_2^f\leftthreetimes G_b$ can be associated with a EF1 topological
order with $G_f=Z_2^f\leftthreetimes \hat G_b$, which may leads to an
understanding of EF2 topological orders in terms of simpler EF1 topological
orders.  
(7) We find
that all EF topological orders correspond to gauged 3+1D fermionic symmetry
protected topological (SPT) orders with a finite unitary symmetry group. Our
results can also be viewed as a classification of the corresponding 3+1D
fermionic SPT orders.  (8) We further propose that the general classification
of 3+1D topological orders with finite unitary symmetries for bosonic and
fermionic systems can be obtained by gauging or partially gauging the finite
symmetry group of 3+1D SPT phases of bosonic and fermionic systems.

\end{abstract}

\maketitle

{\small \setcounter{tocdepth}{2} \tableofcontents }

\section{Introduction}

In \Ref{LW170404221}, we classified the so called all-boson (AB) 3+1D
topological orders -- the 3+1D topological orders whose emergent pointlike
excitations are all bosonic.  We found that \emph{All 3+1D AB topological
orders are classified by pointed unitary fusion
2-categories with trivial 1-morphisms, which are one-to-one labeled by a pair
$(G,\om_4)$ up to group automorphisms, where $G$ is a finite group and $\om_4$
its group 4-cohomology class: $\om_4 \in H^4(G;\R/\Z)$}. 

In this paper, we classify 3+1D topological orders with emergent
fermionic pointlike excitations, which will be called EF topological orders.
The results in \Ref{LW170404221} and in this paper classify all  3+1D
topological orders in bosonic systems.  This result in turn leads to a
classification of 3+1D topological orders with finite unitary symmetry for
bosonic and fermionic systems.  In addition, we argue that all 3+1D bosonic
topological orders always have gappable boundary.

The pointlike excitations and the stringlike excitations in 3+1D bosonic
topological orders can fuse and braid, and their fusion and braiding must form
a self-consistent structure.  In particular,  the self-consistent structure
must satisfy \myfrm{ \textbf{The principle of remote detectability:} In an
anomaly-free topological order, every topological excitation can be detected by
other topological excitations via some remote operations.  If every topological
excitation can be detected by other topological excitations via some remote
operations, then the topological order is anomaly-free.}  Here ``anomaly-free''
means realizable by a local bosonic lattice model in the same dimension
\cite{W1313}.  The  remote detectability condition is also the anomaly-free
condition.

Since the remote detection is done by braiding, the self consistency of fusion
and braiding, plus the remote detectability can totally fix the structure of
pointlike and stringlike excitations. Those  structures in turn classify the
3+1D EF topological orders.  

\section{Summary of results}

\subsection{Emergence of a group $G_f$}

In particular, we show that the pointlike excitations are described by a
symmetric fusion category $\sRep(G_f)$.  In other words, each type of pointlike
excitations correspond to an irreducible representation of a finite group
$G_f$.  The quantum dimension of the excitations is given by the dimension of
the representation.  $G_f$ is a $Z_2^f$ central extension of $G_b$:
\begin{align}
 1 \to Z_2^f \to G_f \stackrel{\pi^f}{\to} G_b\to 1 .
\end{align}
The  excitation is fermionic if $Z_2^f$ is represented non-trivially
in the representation.  Otherwise, the excitation is bosonic.  

\subsection{Unique canonical gapped boundary described by a 
unitary fusion 2-category }
\label{uf2c}

Following a similar approach proposed in \Ref{LW170404221}, in this paper, we
show that all EF topological orders have a unique canonical gapped boundary,
which is described by a unitary fusion 2-category $\sA_b^3$.  Let us describe
such fusion 2-categories in details.  The simple objects of fusion 2-category,
corresponding to the boundary strings, are labeled by $\hat G_b$. Here $\hat
G_b$ is an extension of $G_b$ by $Z_2^m$:
\begin{align}
 1 \to Z_2^m \to \hat G_b  \stackrel{\pi^m}{\to} G_b\to 1 .
\end{align}
The fusion of those boundary strings (the objects) is described by the group
multiplication of $\hat G_b$.  

In the fusion 2-category, there is a 1-morphism of unit quantum dimension that
connects each simple object $g$ to itself.  Such a 1-morphism correspond to a
pointlike topological excitation living on the string $g$. 
But this pointlike excitation is not confined to certain strings; they can move freely on
the boundary and braid among themselves.  The statistics of this pointlike
excitation (the 1-morphism) is fermionic.  So the canonical boundary of a EF
topological order also contains a fermion in addition to the boundary strings.

There is also a 1-morphism of quantum dimension $\sqrt 2$ that connects object
$g$ to object $gm$ where $m$ is the generator of $Z_2^m$.  Physically, it means
that the domain wall between string $g$ and string $gm$ carries a fractional
degrees of freedom of dimension $\sqrt 2$ (\ie like one half of a qubit).
There is no other 1-morphisms.   

In this paper, we show that each EF topological order corresponds to one such
fusion 2-category.  \Ref{ZLW} shows that for each of such fusion 2-categories,
one can construct a bosonic model to realize a EF topological order who has a
boundary described by the fusion 2-category. Thus, the classification of such
unitary fusion 2-categories corresponds to a classification of 3+1D EF
topological orders.

We note that the boundary fermion can form a $p$-wave topological
superconducting chain,\cite{K0131} which is called a Majorana chain. In fact,
two boundary strings labeled by $g$ and $gm$ differ by attaching such a
Majorana chain.  The 1-morphism of quantum dimension $\sqrt 2$ at the domain
wall between the strings $g$ and $gm$ is nothing but the Majorana zero mode at
the end of the Majorana chain.

\begin{figure}[tb] 
\centering \includegraphics[scale=0.8]{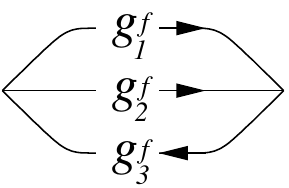} 
\caption{
A string configuration in the bulk described by a triple
$(\chi_{g^f_1},\chi_{g^f_2},[g_3^f])$, where $\chi_{g^f}$ is a conjugacy class in $G_f$
containing $g^f\in G_f$  and the triple satisfy $g_1^f g_2^f=g_3^f$.
}
\label{3strings} 
\end{figure}

\subsection{Emergence of Majorana zero modes}

The above classification of EF topological orders allows
us to divide those  EF topological orders into EF1 topological orders when
$\hat G_b= Z_2^m \times G_b $, and EF2 topological orders when $\hat G_b$ is a
non-trivial $Z_2^m$ extension of $G_b$, described by a group 2-cocycle
$\rho_2(g_b,h_b) \in H^2(G_b,Z_2^m)$.  In the following, we will describe how
to directly measure the group 2-cocycle $\rho_2$  via the Majorana zero modes
carried by the intersections of three strings.  

Consider a \emph{fixed set} of strings labeled by $\chi_{g^f}$ where $\chi_{g^f}$ is a
conjugacy class in $G_f$ that containing $g^f \in G_f$.  Three strings
$\chi_{g^f_1}$, $\chi_{g^f_2}$, and $\chi_{g^f_3}$ can annihilate if $g_1^f g_2^f=g_3^f$.  If
the triple string intersection has a Majorana zero mode, we assign
$\rho^f_2(g_1^f,g_2^f)=-1$.  If the triple string intersection has no Majorana
zero mode, we assign $\rho^f_2(g_1^f,g_2^f)=1$.  (When $G_f$ is Abelian, the
apearance of Majorana zero modes can be determined by the 2-fold topological
degeneracy for the configuration Fig. \ref{3strings}.)  $\rho^f_2(g_1^f,g_2^f)$
only depends on the conjugacy classes of $g_1^f$, $g_2^f$, and $g_3^f$.  Thus
$\rho^f_2$ satisfies
\begin{align}
 \rho^f_2(g_1^f,g_2^f) = \rho^f_2(h_1 g_1^f h_1^{-1},h_2 g_2^f h_2^{-1}),\ \ \ \
h_1,h_2 \in G_f.
\end{align}
It turns out that $\rho^f_2(g_1^f,g_2^f)$ is actually a function on $G_b$, \ie
it has a form
\begin{align}
 \rho^f_2(g_1^f,g_2^f) = \t\rho_2[\pi^f(g_1^f),\pi^f(g_2^f)] .
\end{align}
$\t\rho_2$ in the above is cohomologically equivalent to $\rho_2$ that
describes the extension $\hat G_b$; in other words, we measured $\rho_2$ up
to coboundaries. If the measured
$\rho_2$ is trivial in $H^2(G_b,Z_2^m)$, the corresponding bulk
topological order is a EF1 topological order.  If the measured $\rho_2$ is a
non-trivial cocycle, we get a EF2 topological order.


\subsection{Classification of EF1 topological order by a class of pointed
unitary fusion 2-category}

For an EF1 topological order, the unitary fusion 2-category that describe its
canonical boundary can be simplified, since we can treat the Majorana chain as
a trivial string when $\hat G_b= Z_2^m \times G_b $.  The  simplified unitary
fusion 2-category $\bar \sA_b^3$ has simple objects labeled by $G_b$ and an
1-morphism of unit quantum dimension that connects each simple object to
itself.  There is no other morphisms.  We studied this case thoroughly, and
showed that $\bar \sA_b^3$ are classified by data $(G_b,e_2,n_3,\nu_4)$, where
$G_b=G_f/Z_2^f$, $e_2\in H^2(G_b,\Z_2)$ the 2-cocycle determining the extension
$Z_2^f\to G_f\to G_b$, $n_3\in H^3(G_b,\Z_2)$, and $\nu_4$ is a 4-cochain in
$C^4(G_b,U(1))$ satisfying 
\begin{align}
      \label{dom4n3}
      \dd \nu_4=(-)^{n_3\hcup{1} n_3+e_2\hcup{} n_3}.
\end{align}
The above data $(G_b,e_2,n_3,\nu_4)$ classify the EF1 topological orders.  This
result is closely related to a partial classification of fermionic
symmetry-protected topological (SPT) phases \cite{GW1441}, where a similar
twisted cocycle condition \eqn{dom4n3}  was first obtained (without the
$e_2\hcup{} n_3$ term).

Given a unitary fusion 2-categories $\sA_b^3$ in Section \ref{uf2c}, we can
obtain a pointed unitary fusion 2-categories $\bar \sA_b^3$ by ignoring the
quantum-dimension-$\sqrt 2$ 1-morphisms.
Thus there is a map from the unitary fusion 2-categories $\sA_b^3$ to the
pointed unitary fusion 2-categories $\bar \sA_b^3$.  In other words, there is a
map from EF topological orders to EF1 topological orders.  This relation allows
us to construct a generic EF topological order from a EF1
topological order.

\subsection{A general classification of 3+1D topological orders
with finite unitary symmetry for bosonic and fermionic systems}

With the above classification results, we further propose that the general
classification of 3+1D topological orders with symmetries can be obtained by
gauging 3+1D SPT phases. Partially gauging a SPT phase leads to a phase with
both topological order and symmetry, namely a symmetry-enriched topological
(SET) phase, while fully gauging the symmetry leads to an intrinsic topological
order. The phases in the same gauging sequence share the same classification
data, as the starting SPT phase and the ending topological order coincide in
their classification.

\subsection{ The line of arguments }

The key result of this paper, the classification of 3+1D EF topological orders
is obtained via the following line of arguments.  We first show that condensing
all the bosonic pointlike excitation in a 3+1D EF topological order always give
rise to a unique $Z_2^f$ topological order (see Section \ref{cond}). We
then show that there is a gapped domain wall between the EF and the $Z_2^f$
topological orders (see Section \ref{sec:dw}), and there is a gapped boundary
for the EF topological order (see Section \ref{bndry}).   This
allows us to show that all 3+1D EF topological orders have gapped boundary. The domain wall and the
boundary are described by unitary fusion 2-categories.  This leads to a
classification of 3+1D EF topological orders in terms of a subclass of unitary
fusion 2-categories.

\section{Condensing all the bosonic pointlike  excitations
to obtain a $Z_2^f$ topological order}

\label{cond}

\begin{figure}[tb] 
\centering \includegraphics[scale=0.5]{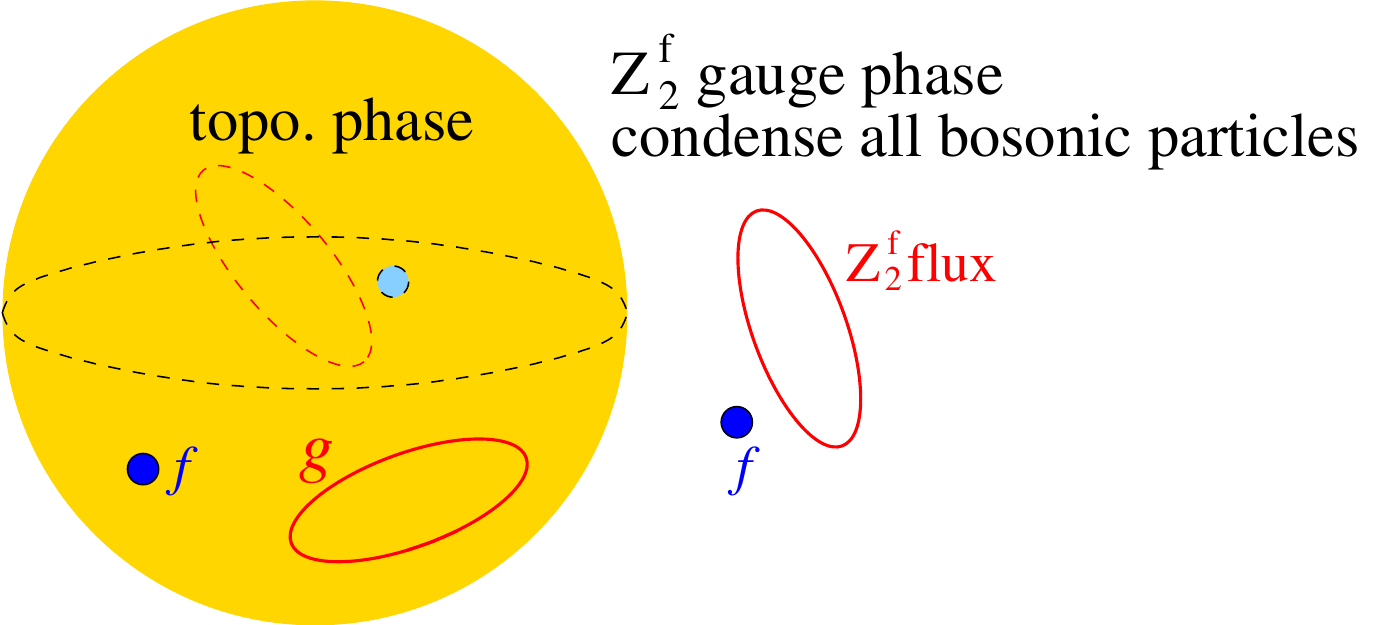} 
\caption{
Condensing all bosonic pointlike excitations in a 3+1D EF topological order
$\sC^4_{EF}$ gives rise to 3+1D $Z_2^f$ topological order $\sC^4_{Z_2^f}$.
$\sC^4_{EF}$ contain a fermionic pointlike excitation $f$, and a stringlike
excitation, $Z_2^f$-flux, which behave like the $\pi$-flux line for the fermion
$f$.  The domain wall $\sA_w^3$ between $\sC^4_{EF}$ and $\sC^4_{Z_2^f}$ contain
strings labeled by elements $g\in G_f$ and only one fermionic particle $f$.
The strings and the fermion have quantum dimension 1.
}
\label{partcondF} 
\end{figure}

Some pointlike excitations in a 3+1D EF topological order are bosons and the
others are fermions.  In this section, we show that, by condensing all the
bosonic pointlike  excitations, we will always ends up with a simple $Z_2^f$
topological order -- a topological order described by 3+1D $Z_2$ gauge theory,
but with a fermionic $Z_2$ charge \cite{LW0510} (see Fig. \ref{partcondF}).  In
the next a few subsections, we will introduce related concepts and pictures
that allow us to obtain such a result.

\subsection{Pointlike excitations and group structure in 3+1D EF topological orders}

The pointlike  excitations in 3+1D EF topological orders are described by SFC.
According to Tannaka duality (see Appendiex \ref{TD}), the SFC give rise to a
group $G_f$ such that the pointlike  excitations are labeled by the irreducible
representations of $G_f$.  In addition, $G_f$ contains a $Z_2$ central
subgroup, denoted by $Z_2^f=\{\one,z\}$.  In each irreducible representations
of $G_f$, $z$ is either represented by $I$ or $-I$ (where $I$ is an identity
matrix). If $z=I$, the corresponding pointlike  excitation is a boson.  We note
that all the bosonic pointlike  excitations are described by irreducible
representations of $G_b$, $\Rep(G_b)$, where $G_b=G_f/Z_2^f$.  If $z=-I$, the
corresponding pointlike  excitation is a fermion.  We denote such SFC by
$\sRep(G_f)$.  We see that each 3+1D EF topological order correspond to a pair
of groups $(G_f,Z_2^f)$ where $Z_2^f$ is the $Z_2$ central subgroup of $G_f$.

\subsection{Stringlike excitations in 3+1D EF topological
orders}\label{stringlike}

The pointlike excitations have trivial mutual statistics among them.  One
cannot use the pointlike excitations to detect other  pointlike
excitations by remote operations.  Thus, based on the principle of remote
detectability, there must stringlike excitations in  3+1D EF topological
orders, so that every  pointlike excitation can be detected by some
stringlike excitations via remote braiding. Similarly, every stringlike
excitation can be detected by some pointlike and/or stringlike excitations
via remote braiding.  We see that the properties of stringlike excitations are
determined by the pointlike topological excitations (\ie $\sRep(G)$) to a
certain degree.  

Let us discuss some basic properties of stringlike excitations.  First, similar
to the particle case, a stringlike excitation $s_i$ can be defined via a trap
Hamiltonian $\Del H_\text{str}(s_i)$ which is non-zero along a loop.  The
ground state subspace of total Hamiltonian $H_0+ \sum_i \Del H_\text{str}(s_i)$
define the fusion space of strings $s_i$ (and particles $p_i$ if we also have
particle traps $\Del H(p_i)$): $\cV(M, p_1,p_2,\cdots,s_1,s_2,\cdots)$.  We
note that such a definition relies on an assumption that all the on-string
excitations are gapped.  We argued that this is always the case
\cite{LW170404221}: 
\myfrm{ A stringlike excitation $s_i$ is called simple if
its fusion space cannot be split by any \emph{non-local} perturbations along
the string (\ie the ground state degeneracy cannot be split by any non-local
perturbations of $\Del H_\text{str}(s_i)$.) } We stress that here we allow
non-local perturbations which are non-zero \emph{only} along the string.  The
motivation to use non-local perturbations is that we want separate out the
degeneracy that is ``distributed'' between strings and particles.  The
degeneracy caused by a single string is regarded as ``accidental'' degeneracy.

For example, in a 3+1D $Z_2$-gauge theory, the $Z_2$-gauge-charge has a mod 2
conservation.  Those $Z_2$-charges can form a many-body state along a large
loop, that spontaneously break the mod 2 conservation which leads to a 2-fold
degeneracy.  We do not want to regard such a string as a non-trivial simple
string.  One way to remove such kinds of string as a non-trivial simple string
is to require the stability against non-local perturbations along a simple
string.  Mathematically, if we allow non-local perturbations as morphisms, the
above string from $Z_2$-charge condensation become a direct sum of two trivial
strings.

The fusion of simple strings may give us non-simple strings
which can be written as a direct sum of simple strings
\begin{align}
 s_i\otimes s_j = \bigoplus_k M^{ij}_k s_k .
\end{align}
Using $M^{ij}_k$ we can also compute the dimension of the fusion space when we
fuse $n$ unlinked loops $s_i$ in the large $n$ limit, which is of order $\sim
d_{s_i}^n$. This allows us define the quantum dimension of the $s_i$ string.

Strings (when they are simple contractable loops $S^1$) can also shrink to a point
and become pointlike excitations:
\begin{align}
\label{shrink}
 s_i \to \bigoplus_j L^i_j p_j.
\end{align}
If the shrinking of a string does not contain $\one$, then we say that the
string is not \emph{pure}.  Such a non-pure string can be viewed as a bound
state of pure string with some topological pointlike excitations.  

In fact, not only strings have shrinking operation, particles also have
shrinking operation.  We note that a zero-dimension sphere $S^0$ is \emph{two}
points, which may correspond to a pair of particles $(p_1,p_2)$.  Thus in
various dimensions $n$, we may have excitations described by $S^d$. For
$d=0,1,2,\cdots$, they correspond to a pair of particles $(p_1,p_2)$, a loop
excitation $s$, a spherical membrane excitation $m$, \etc.  Those
excitations are pure if their shrinking contains $\one$.  For example an $S^0$
excitation $(p_1,p_2)$ is pure iff $p_2$ is the anti particle of $p_1$.

There is a well known result that $p$ is simple iff the shrinking of $p$ and
$\bar p$ (\ie the fusion of $p$ and $\bar p$) contains only a single trivial
particle $\one$.  In this case, we also say that the corresponding pure $S^0$
excitation $(p,\bar p)$ is simple.  Similarly, we believe that \myfrm{A string
$s$ is not simple if the shrinking of $s$ contains more than one trivial
particles $\one$: $s \to  n \one \oplus \cdots$, $n>1$.}

In this paper, we will refer to the number of simple stringlike excitations as
the number of types.  We will refer to the number of pure simple stringlike
excitations as number of pure types.  A string $s$ with quantum dimension $1$
is always simple. Such a string is invertible or pointed, \ie there exists
another string $s'$ such that
\begin{align}
 s\otimes s'=s'\otimes s=\one.
\end{align}
For a more detailed
discussion about stringlike excitations and their related membrane operators,
see \Ref{LW170404221}.

\subsection{Dimension reduction of generic topological orders}

\label{dimred}

\begin{figure}[tb] 
\centering \includegraphics[scale=0.6]{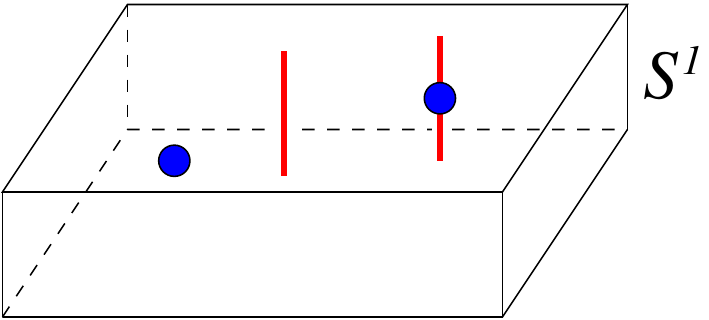} 
\caption{(Color online)
The dimension reduction of 3D space $M^2\times S^1$ to 2D space $M^2$. The top
and the bottom surfaces are identified and the vertical direction is the
compactified $S^1$ direction.  A 3D pointlike excitation (the blue dot)
becomes an anyon particle in 2D.  A 3D stringlike excitation wrapping around
$S^1$ (the red line) also becomes an anyon particle in 2D.
}
\label{3D2D} 
\end{figure}

We can reduce a $3+1$D topological order $\sC^{4}$ on space-time $M^3\times
S^1$ to $2+1$D topological orders on space-time $M^3$ by making the circle
$S^1$ small (see Fig.\,\ref{3D2D}) \cite{MW1514,WW1454}.  In this limit, the $3+1$D topological
order $\sC^{d+1}$ can be viewed as several $2+1$D topological orders
$\sC^{3}_i$, $i=1,2,\cdots,N^\text{sec}_1$ which happen to have degenerate
ground state energy.  We denote such a dimensional reduction process by
\begin{equation}
\label{C3C2}
\sC^{4} = \bigoplus_{i=1}^{N^\text{sec}_1} \sC^{3}_i,
\end{equation}
where $N^\text{sec}_1$ is the number of sectors 
produced by the dimensional reduction.

We note that the different sectors come from the different holonomy of moving
pointlike excitations around the $S^1$ (see Fig.\,\ref{3D2D}).  So the
dimension reduction always contain a sector where the holonomy of moving any
pointlike excitations around the $S^1$ is trivial.  Such a sector will be
called the untwisted sector.  

In the untwisted sector, there are three kinds of anyons.  The first kind of
anyons correspond to the 3+1D pointlike excitations.  The second kind of
anyons correspond to the 3+1D pure stringlike excitations wrapping around the
compactified $S^1$.  The third kind of anyons are bound states of the first two
kinds (see Fig.\,\ref{3D2D}).

\begin{figure}[tb] 
\centering \includegraphics[scale=0.4]{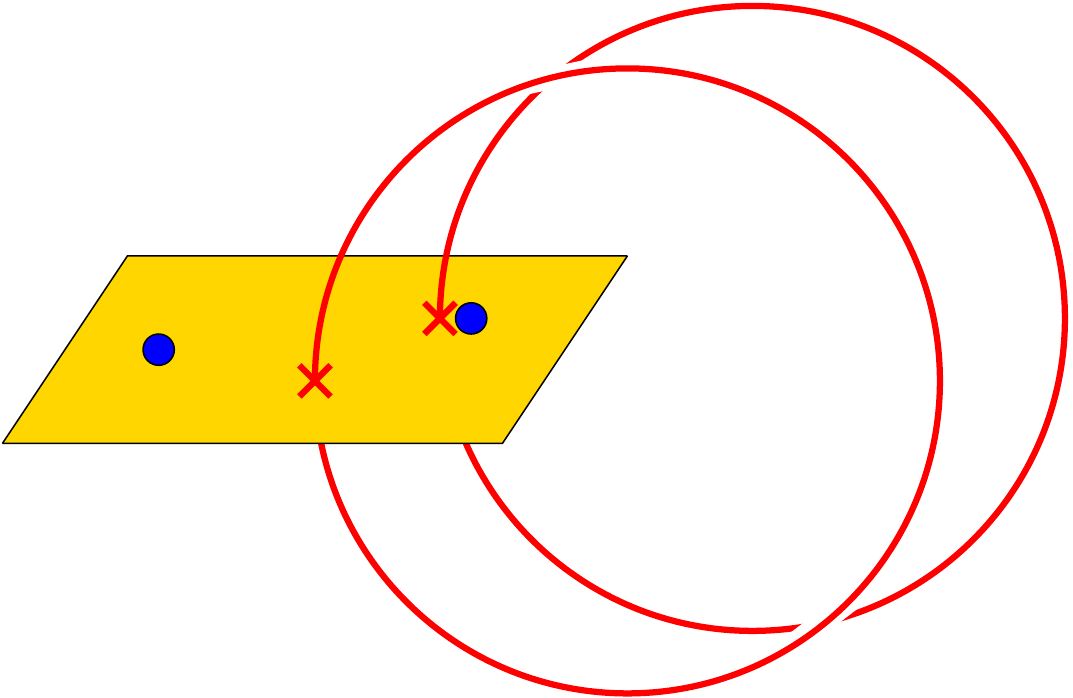} 
\caption{(Color online)
The untwisted sector in the dimension reduction can be
realized directly on a 2D sub-manifold in 3D space without compactification.
}
\label{3D2Dsub} 
\end{figure}

We like to point out that the untwisted sector in the dimension reduction can
even be realized directly in 3D space without compactification. Consider a 2D
sub-manifold in the 3D space (see Fig.\,\ref{3D2Dsub}), and put the 3D pointlike
excitations on the 2D sub-manifold.  We can have a loop of string across the 2D
sub-manifold which can be viewed as an effective pointlike excitation on the 2D
sub-manifold. We can also have a bound state of the above two types of effective
pointlike excitations on the 2D sub-manifold.  Those effective pointlike
excitations on the 2D sub-manifold can fuse and braid just like the anyons in
2+1D.  The principle of remote detectability requires those effective pointlike
excitations to form a unitary modular tensor category (UMTC).  When we perform dimension reduction, the above UMTC
becomes the untwisted sector of the dimension reduced 2+1D topological order.

Since the dimension reduced 2+1D topological orders must be anomaly-free, they
must be described by UMTCs.  Since the untwisted sector
always contains $\sRep(G_f)$, we conclude that \myfrm{The untwisted sector of a
dimension reduced 3+1D EF topological order is a modular extension of
$\sRep(G_f)$.}

\subsection{Untwisted sector of dimension reduction is the 2+1D Drinfeld center}

In the following we will show a stronger result, for the dimension reduction of
generic 3+1D topological orders.  Let the symmetric fusion category formed by
the pointlike excitations be $\sE$, $\sE=\Rep(G)$ or $\sE=\sRep(G_f)$ for AB or
EF cases respectively:  \myfrm{The untwisted sector $\sC^3_\text{untw}$ of
dimension reduction of a generic 3+1D topological orders must be the 2+1D
topological order described by Drinfeld center of $\sE$:
$\sC^3_\text{untw}=Z(\sE)$.} Note that Drinfeld center $Z(\sE)$ is the minimal
modular extension of $\sE$.

First, let us recall the definition of Drinfeld center. The Drinfeld center
$Z(\sA)$ of a fusion category $\sA$, is a braided fusion category, whose
objects are pairs $(A,b_{A,-})$, where $A$ is an object in $\sA$, $b_{A,-}$ is
a set of isomorphisms $b_{A,X}:A\otimes X\cong X\otimes A,\forall X\in \sA$.
The isomorphisms $b_{A,X}$ is just the collection of unitary operators that connects the
fusion spaces $\cdots \otimes A\otimes X \otimes \cdots$ and $\cdots \otimes
X\otimes A \otimes \cdots$ for different backgrounds. They satisfy some self consistent conditions such as
the hexagon equation:
\begin{align}
  b_{A,Y} b_{A,X} = b_{A,X\otimes Y},
\end{align}
where we omitted the associativity constraints (or F-matrices) of $\cA$ for
simplicity (otherwise there are in addition three F-matrices involved, in total
six terms, hence the name hexagon).
$b_{A,X}$ is called a half braiding.

Physically, we may view the objects in $\sA$ as the pointlike topological
excitations living on the boundary of a 2+1D topological order.  In general, a
boundary excitation trapped by a potential on the boundary cannot be lifted
into the bulk.  Physically, this mean that as moving the trapping potential into
the bulk, the ground state subspace will be joined by some high energy
eigenstates to form a new ground state subspace.  But we may choose the boundary
trapping potential very carefully, so that ground state subspace is formed by
accidentally degenerate boundary excitations.  In this case, we say that the
excitation trapped by the boundary potential is a direct sum of those boundary
excitations.  Such an excitation correspond to a composite object in the fusion
category $\sA$.  Now the question is that which composite object (or  direct
sum of boundary excitations) can be lifted into the bulk (\ie the ground state
subspace only rotates by unitary transformation as we move the trapping
potential into the bulk)?

\begin{figure}[tb]
  \centering
  \includegraphics[scale=0.5]{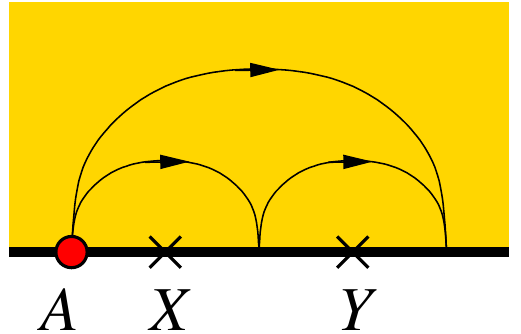}
  \caption{(Color online)
If a (composite) boundary excitations can be lifted in to the bulk, its half
braiding with other boundary excitations must satisfy some self consistent
conditions. The above illustrates the hexagon equation $ b_{A,Y} b_{A,X} =
b_{A,X\otimes Y}$.
}
  \label{hBrd}
\end{figure}

We try to answer this question by exchanging a composite object $A$ in $\sA$
with an arbitrary boundary excitation $X$ and study the unitary transformation
$b_{A,X}$ induced by such an exchange.  If $A$ can be lifted into the bulk,
this $b_{A,X}$ can be interpreted as coming from the half braiding (see Fig.
\ref{hBrd}). There are self consistent conditions from those half braidings.
If we find a composite object $A$ whose half braidings satisfy those consistent
conditions, we believe that the object $A$ can be lifted into the bulk.

However, there is an additional subtlety: even when we require the ground state
subspace only rotates by unitary transformation as we move the trapping
potential into the bulk, there are still different ways to move a composite
boundary excitation $A$ into the bulk, which 
lead different pointlike excitations in the bulk.  Those different bulk
excitations can be distinguished by their different half braiding properties
with all the boundary excitations $X$.  We assume that all the bulk excitations can
be obtained this way.  Therefore, the bulk excitations are given by 
pairs $(A,b_{A,-})$, which correspond to the objects in the Drinfeld center
$Z(\sA)$. 

Mathematically, the morphisms of $Z(\sE)$ between the pairs
$(A,b_{A,-}),(B,b_{B,-})$ is a subset of morphisms between $A,B$, such that
they commute with the half braidings $b_{A,-},b_{B,-}$. Two pairs
$(A,b_{A,-}),(B,b_{B,-})$ are equivalent if there is an isomorphism in $Z(\sE)$ between
them, namely there is an isomorphism, a collection of unitary operators between the fusion
spaces $\cdots\otimes A\otimes \cdots,\cdots\otimes B\otimes \cdots$ that commutes with the
half braidings $b_{A,-},b_{B,-}$.
The fusion and braiding of $(A,b_{A,-})$'s is given by
\begin{align}
  (A,b_{A,-})\otimes (B,b_{B,-})&=(A\otimes B,
  (b_{A,-}\otimes\id_B)(\id_A\otimes b_{B,-}),\nonumber\\
  c_{(A,b_{A,-}), (B,b_{B,-})}&=b_{A,B}.
\end{align}
In other words, to half-braid $A\otimes B$ with $X$, one just half-braids $B$
and $A$ successively with $X$, and the braiding between $ (A,b_{A,-})$  and
$(B,b_{B,-})$ is nothing but the half braiding.

$\sC^3_\text{untw}=Z(\sE)$ is the consequence that the strings in the untwisted
sectors are in fact shrinkable. From the effective theory point of view, we can
shrink a string $s$ (including bound states of particles with strings, in
particular, pointlike excitations viewed as bound states with the trivial
string) to a pointlike excitation $p^\text{shr}_s$ in $\sE$
\begin{align}
\label{stop}
  s\to p^\text{shr}_s=p_1\oplus p_2\oplus\dots, \quad p_1,p_2,\dots\in\sE
\end{align}
So if we only consider fusion, the particles $s,\ p$ in the dimension reduced
untwisted sector $\sC^3_\text{untw}$ can all be viewed as the particles in
$\sE$, regardless if they come from the 3D particles or 3D strings. In
particular, the particles from the 3+1D strings $s$ can be viewed as composite
particles in $\sE$ (see \eqn{stop}).
Next we consider the braidings of them.

In the untwisted sector, the braiding between strings $s,s'$, denoted by
$c_{s,s'}$, requires string $s'$
moving through string $s$, which prohibits shrinking string $s$.
However,
there is no harm to consider the shrinking if we focus on only the initial and
end states of the braiding process. 

\begin{figure}[tb]
  \centering
  \includegraphics[scale=0.5]{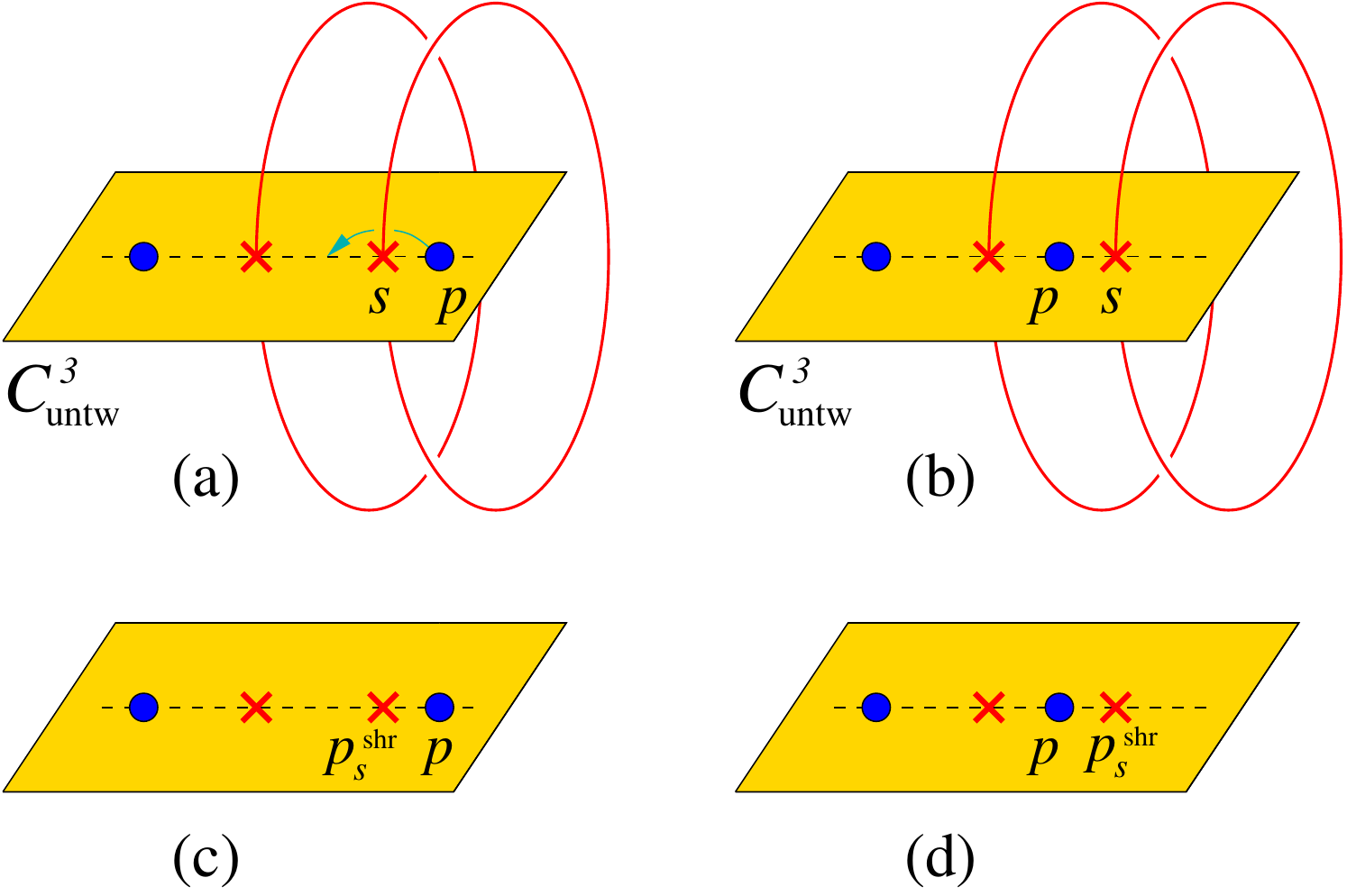}
  \caption{(Color online)
  From (a) to (b) is the braiding $c_{s,p}$ in the untwisted sector.
(c)(d) are obtained from (a)(b) by shrinking strings. Shrinking thus induces a
``half-braiding'' isomorphism $c^\text{shr}_{s,p}$ from (c) to (d).}
  \label{fig:hbraid}
\end{figure}

In particular, the braiding between a string $s$ and a particle $p$, induces an
isomorphism between the initial and end states where the string $s$ is shrunk
(see Fig.\,\ref{fig:hbraid})
\begin{align}
  c^\text{shr}_{s,p}: p^\text{shr}_s\otimes p \cong p \otimes p^\text{shr}_s
\end{align}
which is automatically a half-braiding on the particle $p^\text{shr}_s$.
Thus, $(p^\text{shr}_s,c^\text{shr}_{s,-})$, by definition, is an object in
the Drinfeld center $Z(\sE)$.

Shrinking induces a functor
\begin{align}
  \sC^3_\text{untw}&\to Z(\sE)\nonumber\\
  s&\mapsto (p^\text{shr}_s,c^\text{shr}_{s,-})
\end{align}
which is obviously monoidal and braided, \ie, preserves fusion and braiding. It is also fully faithful, namely
bijective on the morphisms. Physically this means that the local 
operators on both sides are the same.
On the left side, morphisms on a string $s$ are operators acting on near (local to)
the string $s$; on the right side, morphisms in the Drinfeld center are
morphisms
on the particle $p^\text{shr}_s$ which commute with the half braiding
$c^\text{shr}_{s,-}$. From the shrinking picture, morphisms on $p^\text{shr}_s$ can be
viewed as the operators acting on both near the string $s$ and the interior
of the string (namely on a disk $D^2$). But in order to commute with $c_{s,p}$
for all $p$, which can be represented by string operators for all $p$ going through the
interior of the string $s$ (this includes all possible string operators,
because string operators for all particles form a basis), we can take only the operators that act trivially on
the interior of the string. Therefore, morphisms on the right side are also
operators acting on only near the string. This establishes that the functor is
fully faithful, thus a braided monoidal embedding functor; in other words,
$\sC^3_\text{untw}$ can be viewed as a full sub-UMTC of $Z(\sE)$. However, $Z(\sE)$ is already a
\emph{minimal} modular extension of $\sE$, which implies that
\begin{align}
  \sC^3_\text{untw}=Z(\sE).
\end{align}

As $Z(\sE)$ is known well, many properties can be easily extracted. For
example, objects in $Z(\sE)$ have the form $(\chi, \rho)$,
where $\chi$ is a conjugacy class, $\rho$ is a representation of the
subgroup that centralizes $\chi$. One then
concludes
 \myfrm{1. A looplike excitation in a 3+1D topological
order always has an integer quantum dimension, which is $|\chi|\dim \rho$.\\
2. Pure strings ($\rho$ trivial) always correspond to conjugacy classes of the group.}

In particular,  for 3+1D EF topological orders, as the fermion number parity
$z$ is in the center of $G_f$, its conjugacy class has only one element. We
have the following corollary, which is used in later discussions
\myfrm{In all 3+1D EF topological orders, there is an invertible pure $Z_2^f$ flux
loop excitation, corresponding to the conjugacy class of fermion number parity $z$.}

\subsection{Condensing all the bosonic pointlike  excitations}

Starting from a 3+1D EF topological order $\sC^4$, we can condense all the
bosonic pointlike  excitations described by $\Rep(G_b)$, to obtain a new 3+1D
EF topological order $\t\sC^4$.
After $\Rep(G_b)$ is condensed, all bosonic pointlike excitations become the
trivial pointlike excitation in $\t\sC^4$ while all fermionic pointlike
excitations become the same fermionic pointlike excitations with quantum
dimension 1. In other words, the pointlike excitations in the new topological order $\t\sC^4$ are described by $\sRep(Z_2^f)$.

What the stringlike excitations in $\t\sC^4$?  Although the pointlike
excitations in $\t\sC^4$ is very simple and can only detect simple strings,
the stringlike excitations can braid among themselves and detect each other.
Thus $\t\sC^4$ might contain complicated stringlike excitations.

However, using the dimension reduction discussed above, the stringlike
excitations are determined by the pointlike excitations described by
$\sE=\sRep(Z_2^f)$.  In particular, the untwisted sector of the dimension
reduction must be the Drinfeld center $Z(\sE)=Z[\sRep(Z_2^f)]$, which is
nothing but the 2+1D $Z_2$-gauge theory.  There are only four types of 2+1D
anyons: two of them correspond to the 3+1D pointlike excitations in
$\sRep(Z_2^f)$ and the other two  correspond to the 3+1D stringlike
excitations.  The fusion rule between the four anyons in the 2+1D $Z_2$-gauge
theory is described by $Z_2\times Z_2$ group.  This leads to the fusion rule
between the loops and the fermion $f$
\begin{align}
f\otimes f &=\one,\ \ \
 f\otimes s_1 = s_2,\ \ \ 
 f\otimes s_2 = s_1,\ \ \ 
\nonumber\\
 s_1 \otimes s_1 &= s_2 \otimes s_2 = \one, \ \ \ 
 s_1 \otimes s_2 = f . 
\end{align} 
The above also implies the
shrinking rule for the loops to be
\begin{align}
 s_1 \to \one,  \ \ \ \
 s_2 \to f.
\end{align}
We also
find that the braiding phases between the fermion $f$ and the two loops $s_i$
are given by $-1$, and the braiding phase between two $s_1$ or two $s_2$'s is
$1$. The braiding phase between $s_1$ and $s_2$ is $-1$. Here the invertible
loop $s_1$ is the just the $Z_2^f$ flux loop $z$. 

We see that $\t\sC^4$ contains only one type of pure simple string $s_1$
which shrinks to a single $\one$. The other loop $s_2$ is the bound state of
$s_1$ and the fermion $f$.  The loop $s_1$ has a trivial two-loop braiding with
itself.

How many 3+1D EF topological orders that have the above properties?  To answer
such a question, we condense the pure string $s_1$ in $\t\sC^4$ to obtain
a topological order $\sD^4$.  Condensing the pure string $s_1$  corresponds to
condensing the corresponding topological boson in the untwisted sector (which
is described by 2+1D $Z_2$-gauge theory), which changes the untwisted sector to
a trivial phase.  So the untwisted sector of dimension reduced $\sD^4$ is
trivial, which implies $\sD^4$ has no nontrivial particlelike and stringlike
excitations.

We can also obtain such a result by noticing that, in $\sD^4$, the fermions and
$s_2$ are confined (due to the nontrivial braiding with $s_1$) and $s_1$ become
the ground state (\ie condensed). Thus $\sD^4$ has no nontrivial bulk
excitations, and must be an invertible topological order. But in 3+1D, all
invertible topological orders are trivial \cite{K1459,KW1458,F1478}.  Thus
$\sD^4$ is a trivial phase.  This means that we can create a boundary of
$\t\sC^4$ by condensing $s_1$ strings.  Such a boundary contains only one
fermionic particle $f$ with a $Z_2$ fusion rule
\begin{align}
 f\otimes f =\one.
\end{align}
So the boundary is described by a so called unitary braided fusion 2-category
that has no non-trivial objects and has only one non-trivial 1-morphism that
corresponds to a fermion with a $Z_2$ fusion.  It is nothing but the SFC
$\sRep(Z_2^f)$, trivially promoted to a 2-category.  Using the principle that
boundary uniquely determines the bulk \cite{KW1458,KW170200673}, we conclude
that all the $\t\sC^4$'s that satisfy the above properties are actually the
same topological order, which is called $Z_2^f$ topological order
$\sC_{Z_2^f}^4$: \myfrm{Condensing all the bosonic pointlike  excitations in
$\Rep(G_b)$ produces an unique 3+1D topological order $\sC_{Z_2^f}^4$.} The
topological order $\sC_{Z_2^f}^4$ was constructed on a cubic lattice
\cite{LW0316}.  It was also called twisted $Z_2$ gauge theory where the $Z_2$
charge is fermionic, and was realized by 3+1D Levin-Wen string-net model
\cite{LW0510}.  $\sC_{Z_2^f}^4$ can also be realized by Walker-Wang model
\cite{KBS1307} or by a 2-cocycle lattice theory \cite{W161201418}.  In this
paper, we will refer to $\sC_{Z_2^f}^4$ as the $Z_2^f$-topological order.

\section{All 3+1D bosonic topological orders have gappable boundary}
\label{allbndry}

It is well known that 2+1D topological orders with a non-zero chiral central
charge $c$ cannot have gapped boundary.  This can be understood from the
induced gravitational Chern-Simons term in the effective action for such kind
of topological orders.  Since there is no gravitational Chern-Simons term in
3+1D.  This might suggest that all 3+1D bosonic topological orders have
gappable boundary. However, such a reasoning is not correct.  In fact, there
are 2+1D topological orders with a zero chiral central charge (\ie with no
gravitational Chern-Simons term) that cannot have gapped boundary.\cite{L1309} 

For a 2+1D topological order described by a unitary modular tensor category
(UMTC) $\sC^3$, if $\sC^3$ has a \emph{condensable algebra}, then we can
condense the bosons in the condensable algebra to obtain another  2+1D
topological order described by a different UMTC $\sD^3$.  Now we like to ask is
there a gapped domain wall between the two topological orders $\sC^3$ and
$\sD^3$?  In fact, we can show that there exist a 1+1D anomalous topological
order (described by unitary fusion category $\sA_w^2$), such that the Drinfeld
center of $\cA^2$ is $\sC^3\boxtimes \bar \sD^3$.  Here $\sC^3\boxtimes \bar
\sD^3$ is the 2+1D topological order formed by stacking two topological orders,
$\sC^3$ and $\bar \sD^3$, where $\bar \sD^3$ is the time reversal conjugate of
$\sD^3$.  This means that it is consistent to view $\cA^2$ as the domain wall
between $\sC^3$ and $\sD^3$.  Then we conjecture that there exist a gapped
domain wall between $\sC^3$ and $\sD^3$ that is described by $\sA_w^2$.

In the last section, we have seen that condensing all the bosonic excitations
described by $\Rep(G_b)$ in a 3+1D EF topological order $\sC_{EF}^4$ give us an
unique 3+1D topological order $\sC_{Z_2^f}^4$.  This result can also be
obtained by noticing that the condensation of $\Rep(G_b)$ is described by a
condensable algebra \cite{Kon14}, and there is only one condensable algebra if
we want to condense all $\Rep(G_b)$.  So there is only one way to condense all
$\Rep(G_b)$ which produce an unique state $\sC_{Z_2^f}^4$.

Such an unique condensation also produces an unique pointed fusion 2-category
$\sA^3_w$, such that the generalized Drinfield center of $\sA^3_w$ is
$\sC_{EF}^4\boxtimes \bar \sC_{Z_2^f}^4$. Thus it is consistent to view
$\sA^3_w$ as the canonical domain wall between $\sC_{EF}^4$ and
$\sC_{Z_2^f}^4$.  This motivate us to conjecture that \emph{there exist a
gapped domain wall between two 3+1D EF topological orders $\sC_{EF}^4$ and
$\sC_{Z_2^f}^4$}.  

There is a physical argument to support the above conjecture.  The particles in
the condensable algebra are all bosons which form a SFC $\Rep(G_b)$.  Those
bosons have a emergent symmetry described by $G_b$.  Since the number of the
particle types in the condensable algebra is finite, that requires the number
of the irreducible representations of the emergent symmetry group is finite.
Thus the emergent symmetry group $G_b$ is finite.  Those bosons only have short
range interaction between them.  So the boson condensed phase of those bosons
are gapped, with possible ground state degeneracy from the spontaneous breaking
of the emergent symmetry $G_b$.  However since the symmetry is emergent, the
symmetry is only approximate in the boson condensed phase.  The symmetry
breaking term is of an order $\ee^{-l/\xi}$ where $l$ is the mean boson
separation in the boson condensed phase and $\xi$ is the correlation length of
local operators in the topological order.  Since $l$ is finite, the ground
state degeneracy is split  by a finite amount of order $\ee^{-l/\xi}$.  Thus
there is no ground state degeneracy in the boson condensed phase.  This boson
condensed phase corresponds to the $\sC_{Z_2^f}^4$ topological order.

The boson condensed state with a small symmetry breaking perturbation is a very
simple state in physics which is well studied. Such a state always allows
gapped boundary.  Therefore, the domain wall between two 3+1D EF topological
orders $\sC_{EF}^4$ and $\sC_{Z_2^f}^4$ can always be gapped.  {In the last
section, we showed that $\sC_{Z_2^f}^4$ topological order can have a gapped
boundary.} This allows us to argue that all 3+1D EF topological orders have
gappable boundary.

Using a similar argument, we can argue that all 3+1D AB topological orders have
gappable boundary.  In fact, the argument is much simpler in this case. Hence
\myfrm{all 3+1D bosonic topological orders have gappable boundary. }

\section{Unique canonical domain walls between 3+1D EF topological orders and
  $Z_2^f$-topological order $\sC_{Z_2^f}^4$}\label{sec:dw}

In this section, we describe the properties of the fusion 2-category $\sA^3_w$
and show that those properties are consistent of viewing $\sA^3_w$ as a domain
wall between $\sC^4_{EF}$ and $\sC_{Z_2^f}^4$.

\subsection{All simple boundary strings and boundary particles have quantum
dimension 1}

After condensing all bosonic particles $\Rep(G_b)$, the only non-trivial
particle on the canonical domain wall is the fermion $f$ with quantum dimension
1.  Such a fermion can be lifted into  one side of the domain wall with the
$Z_2^f$ topological order $\sC_{Z_2^f}^4$.  On the other side of the domain
wall with 3+1D EF topological order $\sC^4$, if we bring the fermions in
$\sRep(G_f)$ to the boundary, it will become a direct sum (\ie accidental
degenerate copies) of several $f$'s. 

What are the stringlike excitations on the domain wall?  On the
$\sC_{Z_2^f}^4$ side of domain wall, there is only one type of pure simple
stringlike excitations -- the $Z_2^f$ flux loop with quantum dimension 1.
Bring such string to the domain wall will give us a $Z_2^f$ flux loop on the
wall.  We can also bring strings in $\sC^4$ to the domain wall.  In general, a
string in $\sC^4$ will become a direct sum of simple boundary strings.

Let us focus on the simple loop excitations on the canonical domain wall.  A
loop excitation shrunk to a point may become a direct sum of pointlike
excitations (see \eqn{shrink})
\begin{align}
 s \to n \one\oplus m f 
\end{align}
where $\one$ and $f$ are the trivial and fermionic pointlike excitations
respectively.  When $n=0$, the string is not pure. Another possibility is that
$n>1$. In this case the string is not simple.  When $m>1$ the string is also
not simple, since when $s$ fuses with an invertible fermion, its shrinking
rule will become
\begin{align}
 s\otimes f \to m \one\oplus n f,
\end{align}
which is not simple.
Therefore, simple loop excitations on the domain wall have three possible
shrinking rules
\begin{align}
 s_b \to  \one, \ \ \  
 s_f \to   f, \ \ \ 
 s_K \to  \one\oplus  f. 
\end{align}

In the following we would like to show, by contradiction, that a simple string like $s_K$ with
quantum dimension $2$ can not exist on the domain wall.

First, the invertible $Z_2^f$ flux loop $z$, exists in both sides, $\sC^4$ and
$\sC_{Z_2^f}^4$, of the domain wall. We are able to braid $z$ around the
domain wall excitations. As $z$ is invertible, such braiding leads to only a
$U(1)$ phases factor, denoted by $\theta(z,-)$. In particular,
$\theta(z,f)=-1$, which is the defining property of $Z_2^f$ flux.

Second, fusing a fermion $f$ to a string $s_K$ which shrinks to $1\oplus f$,
will not change the string, namely $s_K\otimes f=s_K$. Thus,
\begin{align}
  \theta(z,s_K)=\theta(z,s_K\otimes f)=\theta(z,s_K)\theta(z,f)=-\theta(z,s_K),
\end{align}
which is contradictory. Physically, we can use the braiding of $z$ to detect
the fermion number parity on the domain wall, which implies that excitations
without fixed fermion number parity, such as $s_k\to 1\oplus f$, can not be
stable on the domain wall.
Therefore, there is no simple domain-wall string with quantum
dimension $2$.

Thus, a simple loop on the boundary shrinks to a unique particle, $\one$ or
$f$, with quantum dimension 1.  A simple pure loop on the boundary always
shrinks to a single $\one$.  This is an essential property in the following
discussions: \myfrm{All simple pure loops on the domain wall have a quantum
dimension $d=1$, and their fusion is grouplike.} As the non-pure simple loops
are all bound states of $f$ with pure simple loops, we will consider only the
simple pure loops.  For the moment, we denote the group formed by the simple
pure loops on the domain wall under fusion (see Fig.\,\ref{strfuse}), by $H$.

\subsection{Fusion of domain-wall strings recover the group}
\label{simpleloopgroup}

\begin{figure}[tb]
  \centering
  \includegraphics[scale=0.5]{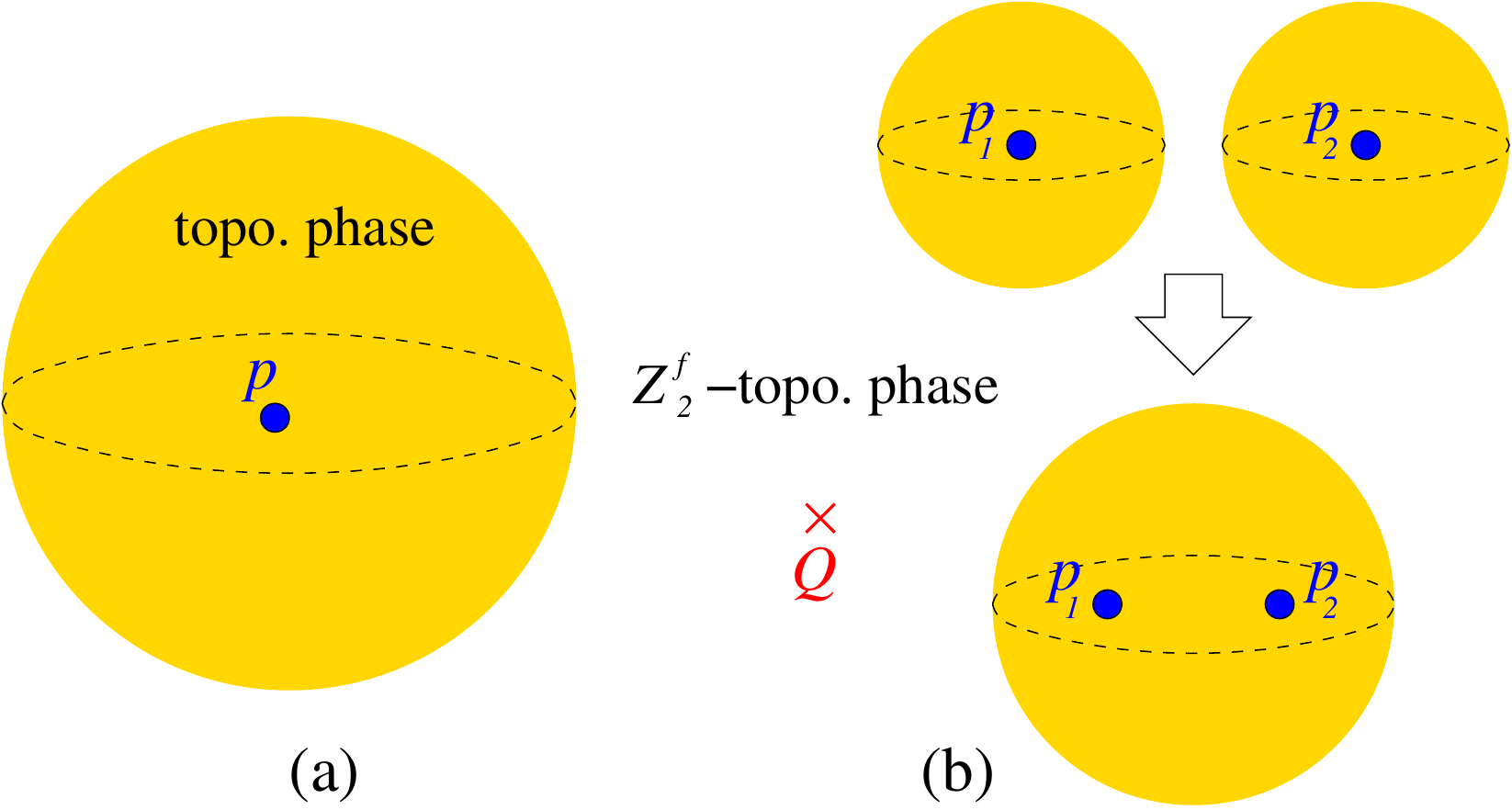}
  \caption{(Color online) (a) The fusion space $F(p)$ for a 3-disk $D^3$ containing only one
particle $p$. (b) Merging two 3-disks to one 3-disk induces an isomorphism
$F(p_1)\otimes_\C F(p_2)\cong F(p_1\otimes p_2)$.}
  \label{fig:fiber}
\end{figure}

The argument in this subsection is almost parallel to those in the AB case
described in \Ref{LW170404221}. There are only a few modifications to address
the fermionic nature. But to be self-contained we include a full argument here.

To apply the Tannaka duality (see Appendiex \ref{TD}), we need a physical
realization of the super fiber functor. Consider a simple topology for the
domain wall: put the 3+1D topological order $\sC^4$ in a 3-disk $D^3$, the
domain wall on $\partial D^3=S^2$, and outside is the condensed phase
$\sC_{Z_2^f}^4$. When there is only a particle $p$ in the 3-disk, a background
particle $Q=\one\oplus f$ in the condensed phase
$\sC_{Z_2^f}^4$,\footnote{Without such background particle, the fusion space
would be 0 if $p$ is a fermion.} with no string and no other particles, we
associate the corresponding fusion space to the particle $p$, and denote this
fusion space by $F(p)$ (see Fig.\,\ref{fig:fiber}). Viewed from very far away,
a 3-disk containing a particle $p$ is like a particle in the condensed phase
$\sC_{Z_2^f}^4$, which has point-like excitations $\sVe=\sRep(Z_2^f)$. When there are two 3-disks,
each containing only one particle, $p_1$ and $p_2$ respectively, the fusion
space is $F(p_1)\otimes_\C F(p_2)$.  Moreover, as adiabatically deforming the
system will not change the fusion space, we can ``merge'' the two 3-disks to
obtain one 3-disk containing one particle $p_1\otimes p_2$. Therefore
$F(p_1)\otimes_\C F(p_2)\cong F(p_1\otimes p_2)$.  Similarly, $F$ also
preserves the braiding of particles. In other words, the assignment $p\to F(p)$
gives rise to a super fiber functor. By Tannaka duality, we can reconstruct a
group $G_f\equiv\Aut(F)$, such that the particles in the bulk $\sC^4$ are
identified with $\sRep(G_f)$.  Our goal is to show that the fusion group $H$ of
the simple loops on the domain wall, is the same as $G_f$.

\begin{figure}[tb]
  \centering
  \includegraphics[scale=0.5]{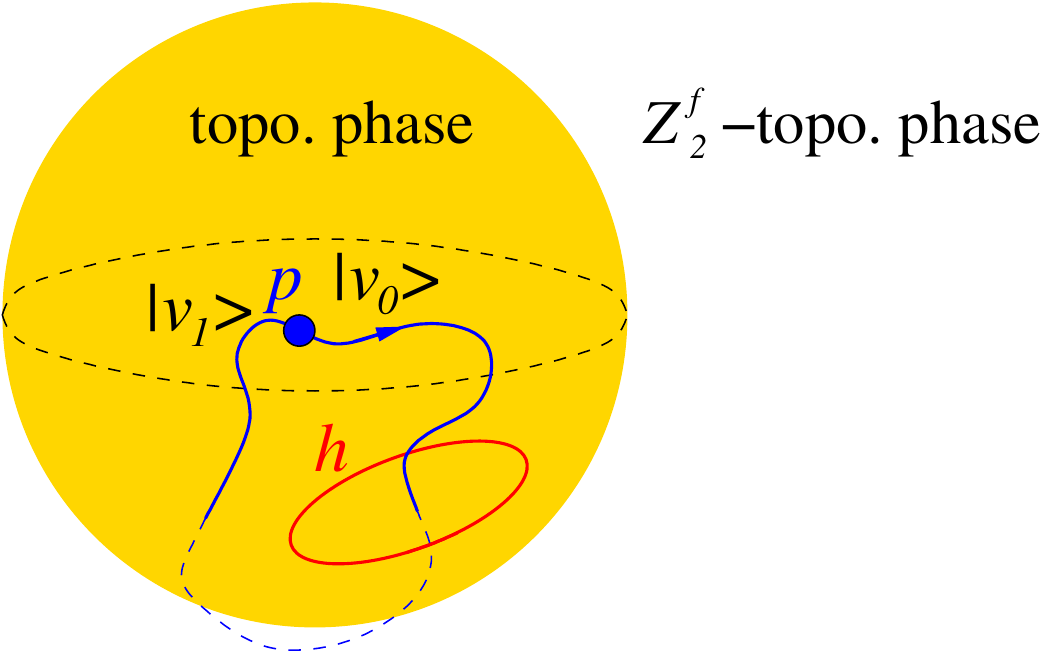}
  \caption{(Color online) Moving a particle (blue) around a loop excitation
(red) on the domain wall. The solid line is in the $\sC^4$ phase. The dashed line
is in the $\sC_{Z_2^f}^4$ phase.}
  \label{halfbraid}
\end{figure}

To do this we consider the process of adiabatically moving a particle $p$
around a pure simple loop $h\in H$ on the domain wall, as shown in
Fig.\,\ref{halfbraid}.  As the pure simple loop is invertible, inserting them
will not change the fusion space. But an initial state $|v_0\>\in F(p)$, after
such an adiabatically moving process, can evolve into a different end state
$|v_1\>\in F(p)$. Thus, braiding $p$ around $h$ induces an invertible (since we
can always move $p$ backwards) linear map on the fusion space $F(p)$,
$\alpha_{p,h}:|v_0\> \mapsto |v_1\>$.

Next, consider that we have two particles $p_1,p_2$ in the bulk. If we braid
them together (fusing them to one particle $p_1\otimes p_2$) around the simple
loop $h$, we obtain the linear map $\alpha_{p_1\otimes p_2,h}$.  If the fusion
of the bulk particles is given by $p_1\otimes p_2 = \bigoplus_i W_i$, we can
split $p_1\otimes p_2$ to the irreducible representations $W_i$, and braid
$W_i$ with $h$. It is easy to see the $\alpha_{p,h}$ maps are automatically
compatible with such splitting (or compatible with the embedding intertwiners
$W_i\to p_1\otimes p_2$); in other words, $ \alpha_{p_1\otimes p_2,h} =
\bigoplus_i \alpha_{W_i,h}.$

But it is also equivalent if we move $p_1,p_2$ one after the other. More
precisely, we can first separate $p_2$ into another 3-disk, braid $p_1$ with
$h$, and then merge $p_2$ back to the original 3-disk.  Thus moving $p_1$ alone
corresponds to the linear map $\alpha_{p_1,h}\otimes_\C \id_{F(p_2)}$.
Similarly, moving $p_2$ alone corresponds to
$\id_{F(p_1)}\otimes_\C\alpha_{p_2,h}$ and in total we have the linear map
${\alpha_{p_1,h}\otimes_\C\alpha_{p_2,h}}$. Therefore, $\alpha_{p_1\otimes
p_2,h}=\alpha_{p_1,h}\otimes_\C \alpha_{p_2,h}$, or using only irreducible
representations,
\begin{align}
  \alpha_{p_1,h}\otimes_\C \alpha_{p_2,h} =
  \bigoplus_i \alpha_{W_i,h}.
\end{align}
These linear maps are compatible with the fusion of bulk
particles.

\begin{figure}[tb] 
\centering \includegraphics[scale=0.5]{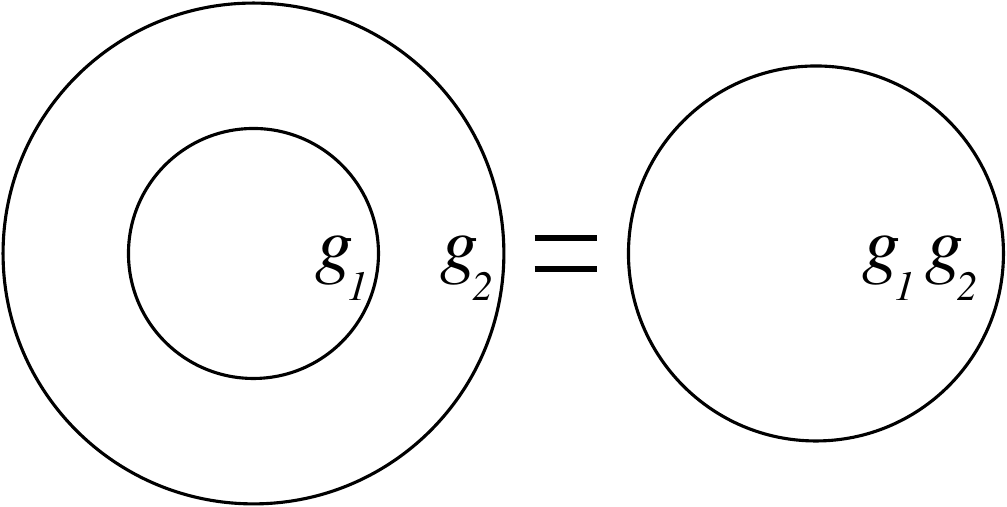} 
\caption{
The fusion of domain wall stringlike excitations $s^\text{bdry}_{g_1} \otimes
s^\text{bdry}_{g_2} = s^\text{bdry}_{g_1 g_2} $ which can be
abbreviated as $g_1
\otimes g_2 = g_1 g_2 $.
}
\label{strfuse} 
\end{figure}

Moreover, the pure simple loop $h$ provides such an invertible linear map
$\alpha_{p,h}$ for \emph{each} particle $p\in\sRep(G_f)$ in $\sC^4$, thus the
set of linear maps
$\varphi(h)\equiv\{\alpha_{p,h}\}$ is an automorphism of the super fiber
functor, $\varphi(h)\in G_f\equiv\Aut(F)$. In
other words, we obtain a map $\varphi$ from the pure simple loops $H$ to $G_f$,
$\varphi:H\to G_f$. It is compatible with the fusion of simple loops, because the
path of braiding around two concentric simple
loops, $g_1,g_2$ (as in Fig.\,\ref{strfuse}), separately, can be continuously deform to the
braiding path around the two loops together, or around their fusion
$g_1\otimes g_2=g_1 g_2$.
This implies that $\varphi(g_1)\varphi(g_2)=\varphi(g_1 g_2)$, namely,
$\varphi$ is a group homomorphism.

Next we show that $\varphi$ is in fact an isomorphism and $H=G_f$.  This is a
consequence of the following principles:
\myfrm{(1) If an excitation has trivial braiding with the condensed
  excitations, it must survive as a de-confined excitation in the
condensed phase.\medskip\\
(2) there is no nontrivial
bulk particle that has trivial half-braiding with all the domain-wall strings.} 
(1) is a general principle for condensations, while (2) is a remote
detectability condition.
  By the folding trick, we
  can regard the domain wall as a boundary of the phase ${\sC^4}\boxtimes
{\sC_{Z_2^f}^4}$.
So we have similar remote detectability condition (2) near the domain wall as
that near a boundary\cite{LW170404221}.

A typical half-braiding path is shown in Fig.\,\ref{halfbraid}, in the sense
that half in $\sC^4$ and half in $\sC_{Z_2^f}^4$.
If $\alpha_{p,h}$ is the identity map, it implies trivial half-braiding between
the particle $p$ in $\sC^4$ and simple loop $h$ on the domain wall.

Now, we are ready to show that $\varphi:H\to G_f$ is an isomorphism:
\begin{enumerate}
  \item $\varphi$ is injective.
    Firstly, the $Z_2^f$ flux loop, denoted by $z$, which is simple, pure,
   invertible and survives in the condensed phase $\sC_{Z_2^f}^4$, must also be a
    pure simple loop on the domain wall. Namely, $Z_2^f\subset H$.

    Consider $\ker\varphi$, namely the pure simple loops
    that induce just identity linear maps on all particles in $\sC^4$. On one
    hand, simple loops in $\ker\varphi$ have trivial
    half-braiding with all particles in $\sC^4$. So they also have trivial
    braiding with the condensed excitations, namely all the bosons in $\sC^4$.
    By (1), they should all
    survive the condensation; in other words, $\ker\varphi$ is at most a
    subset of pure string excitations in $\sC_{Z_2^f}^4$, $\ker\varphi\subset Z_2^f$. 
    On the other hand, the linear map $\alpha_{p,z}$ induced by the $Z_2^f$
    flux loop $z$ is not the identity map on fermions, so $z\notin\ker\varphi$.
    
    Therefore, we see that $\ker\varphi$ must be trivial, which means $\varphi$ is injective.

  \item $\varphi$ is surjective.
    We already showed that $\varphi:H\to G_f$ is injective, so we can view $H$ as
    a subgroup of $G_f$.

    Now consider a special particle in $\sC^4$, which
    carries the representation $\mathrm{Fun}(G_f/H)$,
    linear functions on the right cosets $G_f/H$. More precisely,
    $\mathrm{Fun}(G_f/H)$ consists of all linear functions on $G_f$, $f:G_f\to \C$,
    such that $f(h x)=f(x),$ $\forall h\in H,x\in G_f$ (takes the same value on
    a coset).
    The group action is the usual one on functions,
    ${\rho_{\mathrm{Fun}(G_f/H)}(g):f(x)\mapsto f(g^{-1} x)}$.

    The linear maps $\alpha_{p,h}$ induced by the pure simple loops
    are all actions of group elements in $H$,
    and they are all identity maps on the
    special particle $\mathrm{Fun}(G_f/H)$. In other words, the bulk
    particle $\mathrm{Fun}(G_f/H)$ has trivial half-braiding with all the pure
    domain-wall strings. As a non-pure domain-wall string is just the bound
    state of $f$ with a pure domain-wall string, its half-braiding with
    $\mathrm{Fun}(G_f/H)$ is also trivial. Thus $\mathrm{Fun}(G_f/H)$ has
    trivial half-braiding with all the domain-wall strings. By the remote detectability condition (2), it
    must be the trivial particle carrying the trivial representation.
    In other words, we have $G_f=H$.
\end{enumerate}

To conclude, the pure simple loop excitations on the domain wall, forms a
group under fusion. It is exactly the same group whose representations are
carried by the pointlike excitations in the bulk.

\subsection{Unitary pointed fusion 2-category with a single
invertible fermionic 1-morphism}

In addition to the strings on the domain wall discussed above, the domain wall
also contain a single fermion with quantum dimension 1. 
Summarizing
the above results, we find that \myfrm{a 3+1D EF topological order $\sC_{EF}^4$
has an unique domain wall that connects it to the 3+1D  $Z_2^f$-topological
order $\sC_{Z_2^f}^4$. The domain wall is described by an unitary pointed
fusion 2-category such that for each
object (string) there is only one nontrivial invertible 1-morphism
corresponding to the fermion.}

However, the  domain wall only realize a special subset of unitary pointed
fusion 2-categories with a single invertible fermionic 1-morphism.  The
realized fusion 2-categories, denoted as $\sA_w^3$, must
also have the following property:
\begin{align}
\label{Zstack}
Z(\sA_w^3)=\sC^4_{Z_2^f}\boxtimes (\sC^4_{Z_2^f})_{Z(\sA_w^3)}^\text{cen}.
\end{align}

Here $Z(\sA_w^3)$ is the bulk-center of $\sA_w^3$.
The notion of the bulk-center was introduced in \Ref{KW1458,KWZ1590}
which is a generalization of Drinfeld center to higher categories.
Physically, $Z(\sA_w^3)$ is the
unique 3+1D topological order whose boundary can be $\sA_w^3$.  Since $\sA_w^3$
is a domain wall between $\sC^4_{Z_2^f}$ and $\sC^4_{EF}$, after folding,
$\sA_w^3$ can viewed as the boundary of the stacked topological order
$\sC^4_{Z_2^f}\boxtimes {\sC^4_{EF}}=Z(\sA_w^3)$ (strictly speaking we should
take time-reversal of one component in the folding trick; but here
$\cC^4_{Z_2^f}$ is the same as it time-reversal $\overline{\cC^4_{Z_2^f}}$).  Thus $Z(\sA_w^3)$ contains
$\sC^4_{Z_2^f}$ as a subcategory.  The centralizer of $\sC^4_{Z_2^f}$ in
$Z(\sA_w^3)$ is given by
${\sC^4_{EF}}=(\sC^4_{Z_2^f})_{Z(\sA_w^3)}^\text{cen}$,
and $Z(\sA_w^3)$ happen to be the stacking of $\sC^4_{Z_2^f}$ and its
centralizer: $Z(\sA_w^3)=\sC^4_{Z_2^f}\boxtimes
(\sC^4_{Z_2^f})_{Z(\sA_w^3)}^\text{cen}$.

\section{The unique canonical boundary of 3+1D EF topological orders}
\label{bndry}

Because the fusion 2-category on the domain wall of an EF topological order
$\sC^4_{EF}$ and $Z_2^f$ topological order $\sC^4_{Z_2^f}$ must satisfy the
additional condition \eqref{Zstack}, it is hard to classify such a subset of
fusion 2-categories.  In this section, we are going to construct the unique
canonical boundary for every 3+1D EF topological order, and using the fusion
2-category for such a canonical boundary to classify 3+1D EF topological
orders.

To construct the unique canonical boundary for a 3+1D EF topological order
$\sC^4_{EF}$, we start with the unique canonical domain wall $\sA_w^3$ between
$\sC^4_{EF}$ and $\sC^4_{Z_2^f}$.  We then create a boundary $\sA_{Z_2}^3$ of
$\sC^4_{Z_2^f}$ by condensing the strings in $\sC^4_{Z_2^f}$. As discussed
before, such a boundary is described by 
the SFC
$\sRep(Z_2^f)$, viewed as a unitary fusion 2-category.

\begin{figure}[tb] 
\centering \includegraphics[scale=0.8]{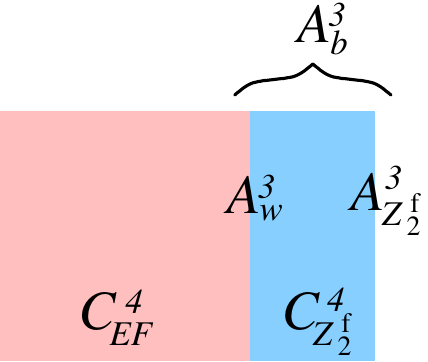} 
\caption{
$\sA_b^3$ is the unique canonical boundary for $\sC^4_{EF}$.  $\sA_b^3$ is
formed by stacking the unique canonical domain wall $\sA_w^3$ between
$\sC^4_{EF}$ and $\sC^4_{Z_2^f}$, and the boundary $\sA^3_{Z_2^f}$ of
$\sC^4_{Z_2^f}$.  Note that $\sA_w^3$ and $\sA^3_{Z_2^f}$ is separated by
$\sC^4_{Z_2^f}$.
}
\label{Ab3} 
\end{figure}

The above construction gives rise to an unique canonical boundary for
$\sC^4_{EF}$ (see Fig.\,\ref{Ab3}):
\begin{align}
 \sA_b^3 = \sA_w^3 \boxtimes_{\sC^4_{Z_2^f}} \sA_{Z_2}^3.
\end{align}

Note that the domain wall $\sA_w^3$ has stringlike excitations labeled by
$G_f$.  But the strings labeled by $Z_2^f \subset G_f$ can move across
$\sC^4_{Z_2^f}$ and then condense on the boundary $\sA_{Z_2}^3$.  So the
stringlike excitations in the whole boundary $\sA_b^3$ are labeled by
$G_f/Z_2^f\equiv G_b$.  All those strings have quantum dimension 1. Their
fusion form the group $G_b$.  The boundary $\sA_b^3$ also contains an invertible
fermion $f$ with quantum dimension 1. Such a pointlike excitation $f$ is
inherited from $\sA_{Z_2}^3$, $\sC^4_{Z_2^f}$, and $\sA_w^3$.  The fermion $f$
can move freely between $\sA_{Z_2}^3$, $\sC^4_{Z_2^f}$, and $\sA_w^3$.  

We like to mention that a ``Majorana chain'' (the 1D invertible fermionic
topological order \cite{K0131}) formed by the boundary fermions may attach to
the strings discussed above.  The Majorana chain is invisible to the braiding
between the stings and particles. But it will double the types of strings.
The end points of such Majorana chains are the quantum-dimension-$\sqrt{2}$
Majorana zero modes. More detailed discussion about this case will be given
later.

Those considerations allow us to obtain the following result (after including
the Majorana chain and doubling the string types): \myfrm{A 3+1D EF topological
order $\sC_{EF}^4$ has an unique boundary $\sA_b^3$.  $\sA_b^3$ is described by
an unitary fusion 2-category whose objects are labeled by $\hat G_b$ which is a
$Z_2^m$ extension of $G_b$, where $Z_2^m$ labels the extra Majorana string.
The fusion of the objects is described by the group multiplication of $\hat
G_b$.  For each object (string) there is one nontrivial invertible 1-morphism
corresponding to the fermion.  There are also quantum-dimension-$\sqrt 2$
1-morphisms (the Majorana zero modes) connecting two objects $g$ and $gm$, with
$g\in \hat G_b$ and $m$ being the generator of $Z_2^m$.} In \Ref{ZLW}, we give
explicit constructions and show that all such unitary fusion 2-categories
correspond to 3+1D EF topological orders.  Classifying such kind of unitary
fusion 2-categories give us a classification of 3+1D EF topological orders.  We
like to remark that $\sA_b^3$ has a form $\sA_b^3 = \sA_w^3
\boxtimes_{\sC^4_{Z_2^f}} \sA_{Z_2}^3$.

The above result allows us to divide the EF topological orders into two groups.
If $\hat G_b =G_b\times Z_2^m$, the corresponding bulk topological orders are
called EF1 topological orders.  The boundary of EF1 topological orders can be
described by a simpler fusion 2-category, since when $\hat G_b =G_b\times
Z_2^m$ we may view the Majorana chain as a trivial string: \myfrm{A 3+1D EF1
topological order $\sC_{EF}^4$ has a unique boundary $\bar \sA_b^3$, which is
described by an pointed unitary fusion 2-category whose objects are labeled by
$G_b$.  The fusion of the objects is described by the group multiplication of 
$G_b$. All 1-morphisms are invertible and fermionic.  There is one nontrivial
1-morphism for each object.}

If $\hat G_b$ is a non-trivial extension of $G_b$ by $Z_2^m$, the corresponding
bulk topological orders are called EF2 topological orders.  In this case, we
cannot view the  Majorana chain as a trivial string.

\section{Classification of EF1 topological orders by pointed unitary fusion
2-categories on the canonical domain wall and boundary}

\subsection{The canonical domain wall}

In this section we will consider the simple case of classification of EF1
topological orders, which is described by the pointed unitary fusion 2-category
$\bar \sA_w^3$ on the domain wall. Such fusion 2-categories are special in the
sense that their objects (corresponding to pure string types) and simple 1-morphisms
are all invertible.
The cases with non-invertible 1-morphisms will be discussed later.

We make the following assumptions:
\begin{enumerate}
  \item The identity (trivial string or trivial particle) related data does not matter.
    The coherence relations involving both the associator/pentagonator and the
    identity related data can be viewed as normalization conditions. We can set
    (by equivalent functors between fusion 2-categories, or physically changing
    the basis or ``gauge'')
    all the identity related data to be trivial, thus the associator
    and the pentagonator are properly normalized.
  \item There are fermions on the strings, but fermions are not confined to the
    strings. Instead, fermions can move freely on the domain wall and even to
    the bulk. As a result, 
    some of the particle related data are fixed by fermionic statistics:
    \begin{align}
      c(f,f)=-1, \quad c(\one,\one)=c(\one,f)=c(f,\one)=1.
    \end{align}
\end{enumerate}
In short, we assume that there is a convenient ``gauge'' choice such that some data
of $\bar\sA_w^3$ are either
normalized or fixed by the fermionic statistics.
\bigskip

\begin{figure}[tb]
  \centering
  \includegraphics[scale=1]{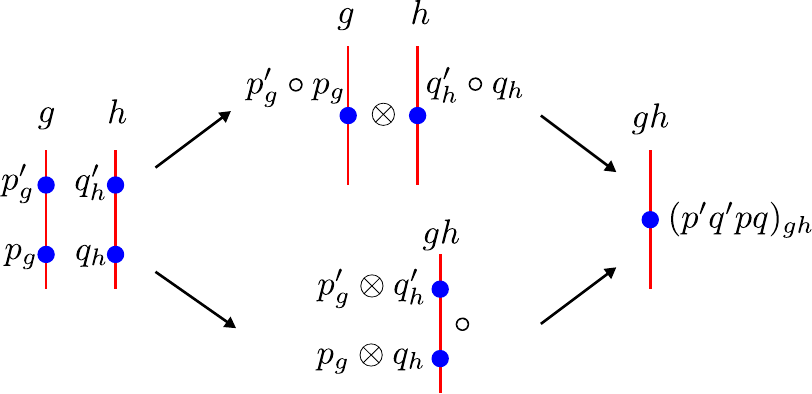}
  \caption{(Color online) The interchange law, corresponding to fusing 4 particles on 2
  strings in different orders. The upper path and the lower path differ by a
$U(1)$ phase $b(p'_g,q'_h,p_g,q_h)$.}
  \label{fig:interchange}
\end{figure}

\textbf{Data}

\begin{enumerate}
  \item Objects (pure string types): $G_f$, a group that has a $Z_2$ central
subgroup. The elements of $G_f$ label the simple pure strings.

  \item 1-morphisms (particles on strings): For any simple pure string labeled
by $g \in G_f$, we have $\Hom(g,g)=\sVe$.  In other words, we have particles
live on a string $g$ which is viewed as a defect between the same type-$g$
string.  $\Hom(g,g)=\sVe$ corresponds to the degenerate subspace or internal
degrees of freedom of the particle.  Here, the particle is in general
composite, which is formed by accidental degeneracy of bosons and fermion, which
in turn gives rise to the super (\ie $Z_2$ graded) vector space $\sVe$.  We
also have $\Hom(g,h)=0$ for $g\neq h \in G_f$.  This means that there is no
1D defect between different simple pure strings.  Simple 1-morphisms are
denoted by $p_g\in \Hom(g,g)$, with a subscript to indicate its string type.
$p$ values in $\{\one,f\}\cong Z_2$, and follows a $Z_2$ fusion rule.

  \item 2-morphisms: linear maps. They correspond to deformation of the particles generated by local operators.

  \item Fusion along strings, denoted by $p_g\circ p_g'$ (composition of
1-morphisms, but in fact is the tensor product in $\sVe$). They follow the
$Z_2$ fusion rule for simple 1-morphisms, $p_g\circ p_g'=(pp')_g$.

  \item Fusion between strings, denoted by $\otimes$, for both objects (given by group multiplication) and
    1-morphisms:
    \begin{align}
      g\otimes h&= gh,\ \ \ \ g,h\in G_f 
\nonumber\\
      p_g\otimes q_h&= (pq)_{gh}.
    \end{align}
As we assume that particles (1-morphisms) can freely move on the domain wall,
the fusion of 1-morphisms along different directions (along or between strings)
should be essentially the same, and independent of the string labels. 

  \item The interchange law, a 2-isomorphism $\tilde b(p'_g,q'_h,p_g,q_h)\in U(1)$
    (see Fig.\,\ref{fig:interchange})
    \begin{align}
      (p'_g\otimes q'_h)\circ(p_g\otimes q_h)\cong (p'_g\circ p_g)\otimes
      (q'_h\circ q_h)
      \label{ic}
    \end{align}
    on $(p'q'pq)_{gh}$.  In our case, the simple strings and simple particles
are all invertible and have quantum dimension 1. Their degenerate subspaces are
always 1-dimensional. Thus the 2-isomorphisms are just $U(1)$ phase factors.

As particles can be freely detached from strings, we expect the above $U(1)$
phase independent of the string labels. Moreover, if we treat the fusion
operations $\circ,\otimes$ as the same one, the difference between the two
sides in \eqref{ic} is just exchanging $q'_h$ and $p_g$. Thus, to be consistent
with fermionic statistics, we assume that
\begin{align}
  \tilde b(p'_g,q'_h,p_g,q_h)=c(q',p).
\end{align}

\begin{figure}[tb]
  \centering
  \includegraphics[scale=0.45]{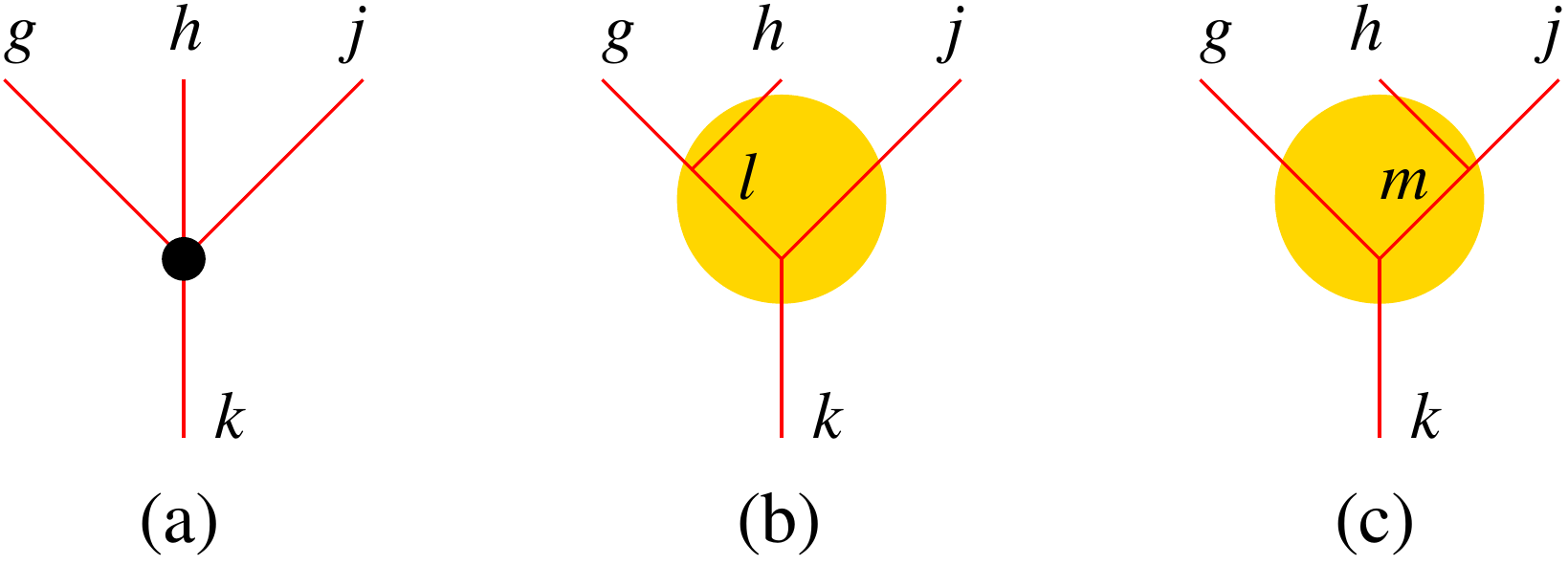}
  \caption{(Color online)
(a) Fusion of strings $g,h,j$ gives rise to a defect between  strings
$g,h,j$ and string $k$.
Two different ways of fusion, (b) and (c), may leads to different defects
whose difference in particles is given by $n_3(g,h,j)$.
}
  \label{strfusion}
\end{figure}

\begin{figure}[tb]
  \centering
  \includegraphics[scale=0.4]{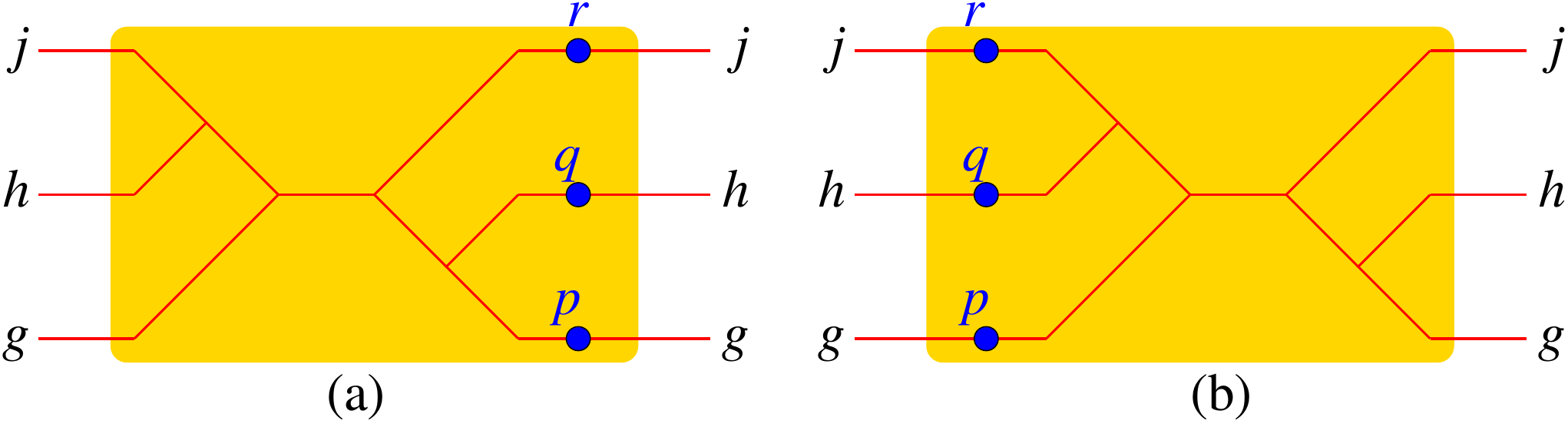}
  \caption{(Color online) 
The two domain-wall states in (a) and (b) may differ by a $U(1)$ phase
$\tilde n_3(p_g,q_h,r_j)$ (see \eqref{an3}).
}
  \label{2morph}
\end{figure}

  \item Associator:
    \begin{itemize}
      \item For $g,h,j\in G_f$, we have a 1-morphism ${n_3(g,h,j): (g\otimes
h)\otimes j\to g\otimes( h\otimes j)}$, valuing in $Z_2=\{\one,f\}$. See  Fig.\,\ref{strfusion}.

      \item We also have a 2-isomorphisms $\tilde n_3(p_g,q_h,r_j) \in U(1)$ to describe
the $U(1)$ phase difference between two different orders to fuse strings and
particles on the strings (see  Fig.\,\ref{2morph}):
	  \begin{align}
	    & n_3(g,h,j)\circ[(p_g\otimes q_h)\otimes
	    r_j]\nonumber\\
	    & \cong[ p_g\otimes(q_h\otimes r_j)]\circ n_3(g,h,j).
	  \end{align}
	  To be consistent with fermionic statistics, we assume that
	  \begin{align}
            \label{an3}
	    \tilde n_3(p_g,q_h,r_j)=c[n_3(g,h,j),pqr].
	  \end{align}
    \end{itemize}

\item Pentagonator: for $g,h,j,k\in G_f$, 2-isomorphism $\nu_4 (g,h,j,k)\in U(1)$:
  \begin{align}
    &[\one_g\otimes n_3(h,j,k)]\circ n_3(g,hj,k)\circ[n_3(g,h,j)\otimes
    \one_k]\nonumber\\
    &\cong n_3(g,h,jk)\circ n_3(gh,j,k)
  \end{align}
\end{enumerate}

\textbf{Axioms}
\begin{enumerate}
  \item $n_3(g,h,j)$ is a normalized 3-cocycle in $H^3(G_f,\Z_2)$.
  \item For $g,h,j,k,l\in G_f$,
    \begin{align}
      &\tilde n_3[n_3(g,h,j)_{ghj},\one_k,\one_l]\tilde n_3[\one_g,n_3(h,j,k)_{hjk},\one_l]\\
      &\ \ \ \ \times \nu_4(h,j,k,l)\nu_4(g,hj,k,l)\nu_4(g,h,j,kl) 
        \nonumber\\
	=&\tilde n_3[\one_g,\one_h,n_3(j,k,l)_{jkl}]
         \nonumber\\
       &\ \ \ \ \times \nu_4(gh,j,k,l)\nu_4(g,h,jk,l)\nu_4(g,h,j,k).\nonumber
    \end{align}
    For convenience, we change the notation a little bit:
    let $n_3(g,h,j)$ value in
    the additive $\Z_2=\{0,1\}$ group instead of the multiplicative
    $Z_2=\{\one,f\}$ (where $n=0$
    corresponds to the trivial boson $\one$,
    and $n=1$ corresponds to the non-trivial
    fermion $f$). Thus,
    \begin{align}
      \tilde n_3[n_3(g,h,j)_{ghj},\one_k,\one_l]&=c[n_3(g,h,j),n_3(ghj,k,l)]\nonumber\\
      &=(-1)^{n_3(g,h,j)n_3(ghj,k,l)},
    \end{align}
    and similarly for other $\tilde n_3$'s. We then have
    \begin{align}
      \label{tw4}
      &\frac{\nu_4(h,j,k,l)\nu_4(g,hj,k,l)\nu_4(g,h,j,kl)}
      {\nu_4(gh,j,k,l)\nu_4(g,h,jk,l)\nu_4(g,h,j,k)}=\\
      &(-1)^{n_3(g,h,j)n_3(ghj,k,l)+n_3(g,hjk,l)n_3(h,j,k)+n_3(g,h,jkl)n_3(j,k,l)}.\nonumber
    \end{align}
    In other words, the 4-cochain $\nu_4(g,h,j,k)$ satisfies
    \begin{align}
      \dd \nu_4=(-)^{\Sq^2 (n_3)},
    \end{align}
    a relation first introduced in \Ref{GW1441},
    where $\Sq^2$ is the Steenrod square and
    $\nu_4$ is normalized.
\end{enumerate}
     Here ``normalized'' means that $n_3(g,h,j)=0$, if any of $g,h,k$ is $\one$
     and $\nu_4(g,h,j,k)=1$, if any of $g,h,j,k$ is $\one$.
\bigskip

We want to point out that by now we considered the consistency conditions
only on the domain wall. There are more constraints when we take into account the
bulk, namely, the bulk-center of the above fusion 2-category should be
$\sC_{EF}^4\boxtimes\sC^4_{Z_2^f}$, in particular the fermion $f$ and the $Z_2^f$ flux $z$
must be liftable and form the 3+1D $Z_2^f$-topological order $\sC^4_{Z_2^f}$.
Unfortunately, we do not have efficient algorithms or theorems to calculate
bulk-centers of fusion 2-categories, which makes it difficult to check
under what extra conditions
the bulk-center of the above fusion 2-category will have the desired form.
Instead we will consider the canonical boundary below.

\subsection{The canonical boundary}
We know that the $\sC^4_{Z_2^f}$
topological order have a gapped boundary by condensing the $Z_2^f$ flux string
$z$. On the gapped boundary there is no string but only one non-trivial particle,
the fermion. Imagine we have the gapped domain wall and gapped boundary as above,
between them is the intermediate  $\sC^4_{Z_2^f}$ phase. Now we squeeze the
intermediate  $\sC^4_{Z_2^f}$ phase to a very thin layer, such that we can view
the composite domain-wall-$\bar\sA_w^3$/$\sC^4_{Z_2^f}$/boundary-$\bar\sA^3_{Z_2^f}$
together as a gapped boundary $\bar\sA_b^3$ of $\sC_{EF}^4$. For such boundary, we only need to
check that in its bulk (the bulk-center), the particles form $\sRep(G_f)$,
which is much easier than checking the bulk-center of the domain wall.
\begin{figure}[tb]
  \centering
  \includegraphics[scale=0.5]{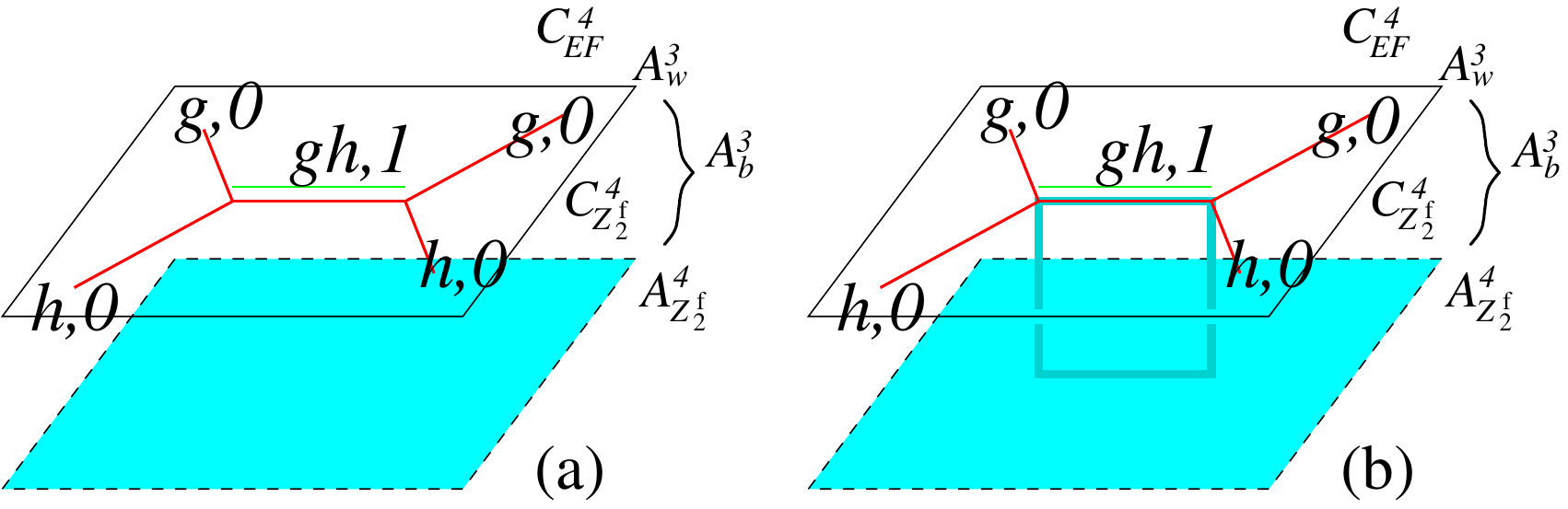}
  \caption{(a) On the domain wall $\bar\sA_w^3$, the strings are labeled by
$(g,\mu)\in G_f$ where $g \in G_b$ and $\mu\in \Z_2^f$.  The fusion of strings
$(g,\mu)$ and $(h,\nu)$ is given by $(g,\mu) \otimes
(h,\nu)=(gh,\mu+\nu+\la_2(g,h))$.  The 2-group-cocycle $\la_2 \in
H^2(G_b,\Z_2^f)$ gives rise to an $Z_2^f$ extension from $G_b$ to $G_f$.  In
the above graph, the string $(g,0)$ is represented by a single line (red) and
the string $(g,1)$ a double line (red,green), where the extra (green) line can
be viewed as the $Z_2^f$ flux line $z$.  (b) Such a $Z_2^f$ flux line can be
canceled by a $Z_2^f$ flux loop $z$ as indicated by the thick rectangular
(blue) loop in the above graph.
}
  \label{strfz}
\end{figure}

\begin{figure}[tb]
  \centering
  \includegraphics[scale=0.5]{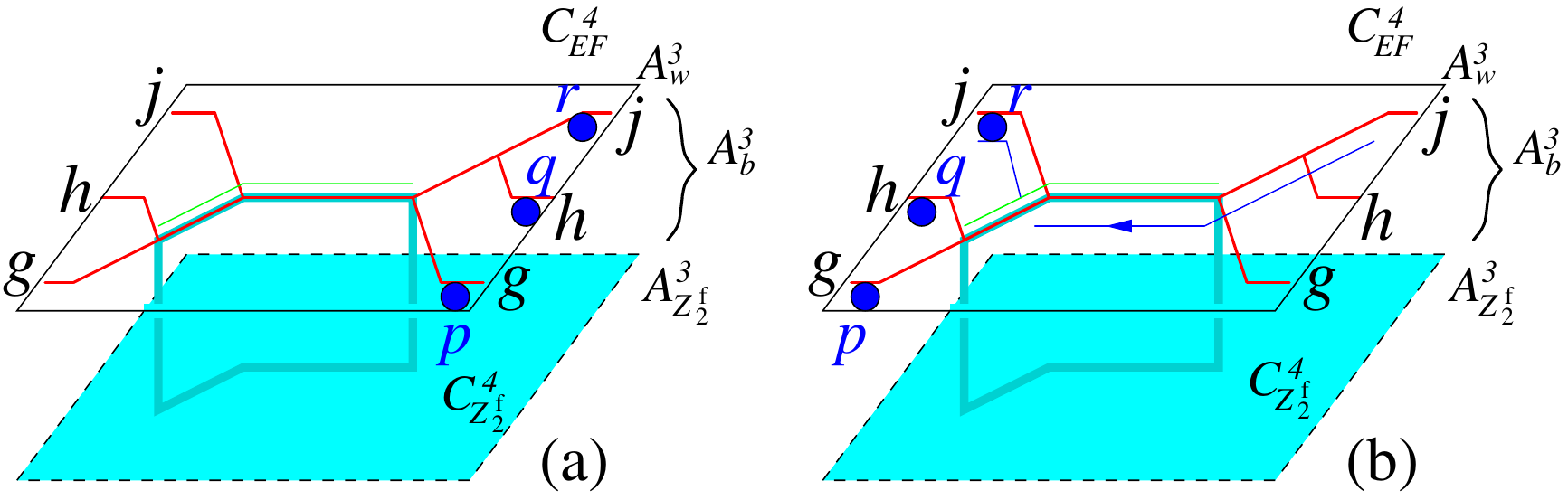}
  \caption{(Color online) 
The two domain-wall states in (a) and (b) may differ by a $U(1)$ phase
$a(p_g,q_h,r_j)$ (see \eqref{an3la2}).
The string label $(g,0)$ on $\bar\sA_w$ is abbreviated to $g$. This figure shows
the case that $e_2(g,h)=
e_2(g,hj)=1,$ ${e_2(gh,j)=e_2(h,j)=0}$.
}
  \label{2morphz}
\end{figure}

The composite boundary is described by a similar fusion 2-category as that for
the domain wall. Most of the data and conditions discussed above apply. We only
list the difference below:
\begin{enumerate}
  \item As the $z$ string condenses, the string types on the boundary are now labeled
    by $G_b=G_f/Z_2^f$. At the same time, the 2-cocycle $e_2(g,h)\in
    H^2(G_b,\Z_2^f)$ coming from the extension $Z_2^f\to G_f\to G_b$ will arise
    in other data (see Fig.\,\ref{strfz}).
  \item When fusing $g,h$ on the composite
    boundary, $e_2(g,h)=1$ indicates that there is a $Z_2^f$ flux loop $z$
    along the fused string $gh$ in the
    intermediate $\sC^4_{Z_2^f}$ phase. As a result, the associator
    $\tilde n_3(p_g,q_h,r_j)$ needs to be modified. Under certain framing convention
    (put the particles slightly below the string in Fig.\,\ref{2morph} and 
slightly into the $\sC^4_{Z_2^f}$ bulk) we find
    that (see Fig.\,\ref{2morphz})
    \begin{align}
      \label{an3la2}
      \tilde n_3(p_g,q_h,r_j)=(-1)^{n_3(g,h,j)(p+q+r)}(-1)^{r e_2(g,h)},
    \end{align}
    where $(-1)^{n_3(g,h,j)(p+q+r)}$ is the fermion statistics (written in the
    additive $\Z_2$ convention) and $(-1)^{r e_2(g,h)}$ is the particle-loop
    statistics coming from $r$ going through the $Z_2^f$ flux loop $z$ along
    $gh$.
  \item $n_3(g,h,j)$ is now a 3-cocycle in $H^3(G_b,\Z_2)$. The condition for 
$\nu_4$ is then modified to
    \begin{align}
      &\frac{\nu_4(h,j,k,l)\nu_4(g,hj,k,l)\nu_4(g,h,j,kl)}
      {\nu_4(gh,j,k,l)\nu_4(g,h,jk,l)\nu_4(g,h,j,k)}=(-1)^{ e_2(g,h)n_3(j,k,l)}\nonumber\\
      & (-1)^{n_3(g,h,j)n_3(ghj,k,l)+n_3(g,hjk,l)n_3(h,j,k)+n_3(g,h,jkl)n_3(j,k,l)}.
    \end{align}
    In other words, the 4-cochain $\nu_4(g,h,j,k)\in C^4(G_b,U(1))$ satisfies
    \begin{align}
      \label{o4gb}
      \dd \nu_4=(-)^{n_3\hcup{1} n_3+ e_2\hcup{} n_3}.
    \end{align}
\end{enumerate}

With these one can check that in the bulk-center bosonic particles form
representations of $G_b$, and fermionic particles form projective
representations of $G_b$ with class described by $ e_2$. Together, particles
form nothing but $\sRep(G_f)$. So the above conditions for the composite
boundary do give rises to a 3+1D EF topological order.  Thus, we have a 
classification of 3+1D EF1 topological orders by $(G_b, e_2,n_3,\nu_4)$, where
$ e_2\in H^2(G_b,\Z_2),n_3\in H^3(G_b,\Z_2),\nu_4\in C^4(G_b,U(1))$ satisfies
\eqref{o4gb}.  The above agrees with the group super-cohomology theory for
fermionic SPTs.  Recently it was found that fermionic SPTs can have ``Majorana
chain layer'' which is beyond the group
super-cohomology\cite{WG170310937,KT1701.08264}. In next subsection we will
show that this ``Majorana chain layer'' also enters in the classification of
topological orders.

For completeness, let us briefly discuss the equivalence relation for the above
data. Firstly, $G_b$ together with $ e_2$ is the same data as the group $G_f$.
Since the particles form $\sRep(G_f)$, by Tannaka duality $(G_b, e_2)$ is fully
determined up to group isomorphisms. However, $(n_3,\nu_4)$ admits more gauge
transformations than co-boundaries: for any 2-cochain $m_2\in C^2(G_b,\Z_2)$
and 3-cochain $\eta_3\in C^3(G_b,U(1))$,
\begin{align}
  n_3&\to n_3+\dd m_2,\\
  \nu_4&\to \nu_4\times \dd \eta_3 \times (-1)^{n_3\hcup{2} \dd
    m_2+m_2\hcup{} m_2+m_2\hcup{1} \dd
  m_2+ e_2\hcup{} m_2}.
\nonumber
\end{align}
give an equivalent solution. Note that $(-1)^{n_3\hcup{2} d m_2+m_2\hcup{}
m_2+m_2\hcup{1} d m_2+ e_2\hcup{} m_2}$ is in general a 4-cochain, and $\dd
\nu_4$ is shifted under such gauge transformation. If we fix $n_3$, namely let
$d m_2=0$, $m_2\in H^2(G_b,\Z_2)$, $\nu_4$ transforms as
\begin{align}
  \nu_4&\to \nu_4\times \dd \eta_3 \times (-1)^{m_2\hcup{} m_2+ e_2\hcup{} m_2},
\end{align}
where $(-1)^{m_2\hcup{} m_2+ e_2\hcup{} m_2}$ is now a 4-cocycle, but may not be the trivial
one. We see that $\nu_4$ is in fact
classified by (forms a torsor over) the group
$H^4(G_b,U(1))/\Gamma$ where $\Gamma$ is the subgroup generated
by $(-1)^{m_2\hcup{} m_2+ e_2\hcup{} m_2}$ for all 2-cocycles $m_2$. Besides the gauge
transformations, different $n_3,\nu_4$ are also equivalent if they can be
related by (outer) group isomorphisms of $G_b$ (which can be followed by gauge
transformations).
To ``add up'' two solutions $(n_3,\nu_4)$ and $(n_3',\nu_4')$, one also
needs to follow a twisted rule,
\begin{align}
  (n_3,\nu_4)+(n_3',\nu_4')=(n_3+n_3',\nu_4\nu_4'(-1)^{n_3\hcup{2}
  n_3'}).
\end{align}

\section{Classification of EF topological orders by unitary fusion
2-categories on the canonical boundary}

\subsection{Define string type using local or non-local unitary transformations?}

\begin{figure}[tb] 
\centering \includegraphics[scale=0.5]{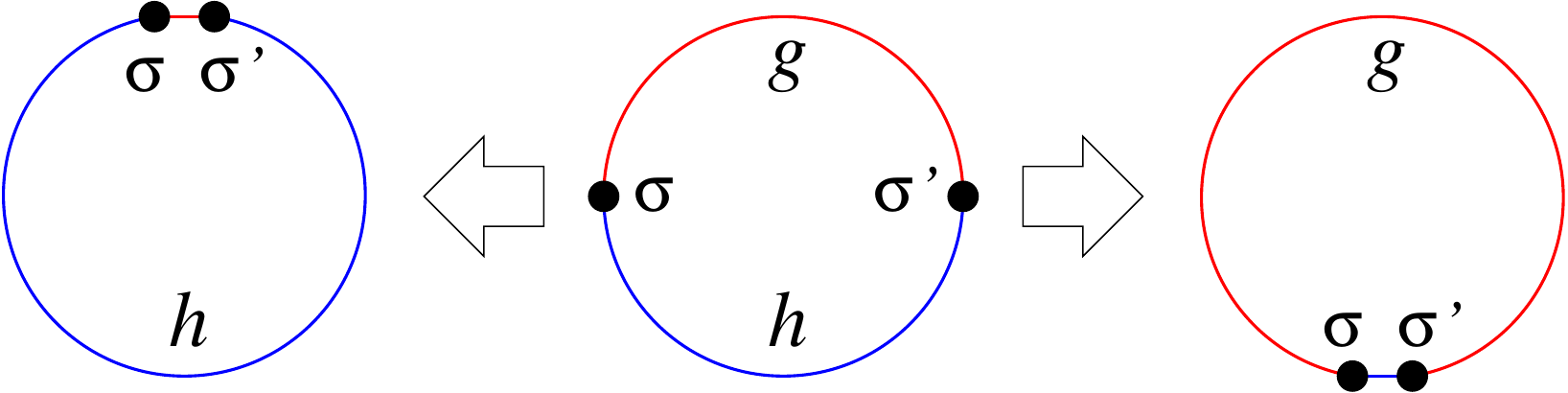} 
\caption{
If two strings $g$ and $h$ can be connected by a domain wall
(\ie an 1-morphism), then under non-local unitary transformations,
strings $g$ and $h$ will be equivalent.
}
\label{gheq} 
\end{figure}

In the above discussions we omitted the possibility that between different
strings there can be 
defects/1-morphisms.  This is a consequence of defining the type of stringlike
excitations up to \emph{non-local} perturbations along the string (see
Sec.~\ref{stringlike}). 
To see this point, let us consider a loop consists of two string segments
labeled by $g,h$ connected by two pointlike defects (i.e.  1-morphisms)
$\sigma\in \Hom (g,h), \sigma' \in \Hom(h,g)$ (see Fig. \ref{gheq}). Under
non-local perturbations, the loop can become a $g$ loop carrying $\sigma\circ
\sigma'\in \Hom(g,g)$, or a $h$ loop carrying $\sigma' \circ\sigma \in
\Hom(h,h)$. Thus $g$ and $h$ will be equivalent under non-local perturbations
along the string.

In the fusion 2-category, the objects/strings and 1-morphisms/point-like defects
are actually defined up to \emph{local} unitary transformations. Moreover, if
there exists an invertible 1-morphism (namely a point-like
defect with quantum dimension 1) between two objects (namely two string
segments), such two objects are equivalent in the fusion 2-category.
Therefore, if some $\sigma\in \Hom (g,h)$ is an
invertible 1-morphism (i.e. its quantum dimension is 1), then $g$ and $h$ are
indeed equivalent as objects in the fusion 2-category, which is consistent with
the non-local perturbation point of view. However, it is possible that no
1-morphism in $\Hom(g,h)$ is invertible, and $g,h$ are not equivalent in the
fusion 2-category.  To include this possibility, we introduce a different
equivalent relation of strings, using local unitary transformations plus
invertible 1-morphisms, which is consistent with that in the fusion
2-category: Two strings defined under local unitary transformations are called
of the same l-type if there is an invertible 1-morphism between them.
The set of l-types will be denoted by $\hat G_b$.
We have already shown that the string types defined via
non-local unitary transformations form a group $G_b$.  Clearly $|\hat G_b| \geq
|G_b|$, and two different l-types may correspond to the same type.




With the expanded string types defined by local unitary transformation, our
arguments in Section \ref{sec:dw} are still valid, which shows that, on the
boundary, closed strings have quantum dimension 1 and form a group under
fusion.  $\hat G_b$ is actually a group that describes the fusion of the
l-types.
Also, using the half braiding with the pointlike
excitation in the bulk (see Section \ref{sec:dw}), we can assign each boundary
string (\ie each l-type) a group element in $G_b$.  Thus there is a group
homomorphism $\hat G_b
\xrightarrow{\pi^m} G_b$. 
If there are non-invertible 1-morphisms between different
l-types, they can together form a closed loop and must be assigned to the same
element in $G_b$.
In fact the string types up to non-local perturbations is just l-types
further up to non-invertible 1-morphisms.
Indeed, $G_b$ is a quotient group of $\hat G_b$ by imposing
equivalent relations via non-invertible 1-morphisms.




\subsection{New string type from Majorana chain}

Next we carefully examine what possible non-invertible 1-morphisms can there be
and their physical meaning.  Since all the l-types of strings 
labeled by $g\in \hat G_b$
have quantum dimension 1 and
form a group under fusion, the 1-morphisms automatically obtain a grading by
this group, namely $p\in \Hom( g, h)$ is graded by $ h g^{-1}$.
As a result of such grading, the total quantum dimension of non-empty
$\Hom( g, h)$ must be the same.  In our previous work discussing AB
topological orders, $\dim \Hom( g, h)=\dim\Hom( g, g)=1$, thus
$\Hom( g, h)$ can only allow one invertible 1-morphism, or be empty; in
this case non-empty $\Hom( g, h)$ just implies $ g= h$.  In
other words in AB topological orders there is no room for non-invertible
1-morphisms on the canonical boundary.  It also means that on the canonical
boundary of AB topological, the string l-types defined using local unitary
transformations plus invertible 1-morphisms and the string types defined using non-local unitary
transformations are the same, \ie $\hat G_b = G_b$.

However, for EF topological orders it is not the case.  Since $\Hom( g,
g)=\sVe$, if $\Hom( g, h)$ is not empty for certain $ g, h$, we
have $\dim \Hom( g, h)=\dim \Hom( g, g)=\dim (\sVe)=2$, which
means that there can be one non-invertible 1-morphism with quantum dimension
$\sqrt2$. In this case $|\hat G_b|> |G_b|$.

We can further fuse a $ g^{-1}$ string to this non-invertible 1-morphism
between $ g, h$, and obtain a non-invertible 1-morphism in $\Hom( g g^{-1}, h
g^{-1})=\Hom(\one, h g^{-1})$. Let such $ h g^{-1}\equiv m$ and denote the
non-invertible 1-morphism by $\sigma_m\in \Hom(\one,m)$. It is easy to see that
for any string $ k$, $\sigma_m \otimes \one_k$ is a non-invertible 1-morphism
in $\Hom( k,m k)$. In fact, such $m$ string generates the kernel of the
projection $\pi^m:\hat G_b\to G_b$.

We find the following
properties of such strings:
\begin{enumerate}
  \item $m$ is a $Z_2$ string, $m^2=\one$. Consider fusing two $\sigma_m$. We
    obtain $\sigma_m\otimes \sigma_m\in\Hom(\one,m^2)$ whose quantum dimension
    is 2. It can only split as the direct sum of two invertible 1-morphisms.
    This implies that the $m^2$ string and $\one$ are equivalent.
  \item $m$ is unique. Suppose that there is another non-invertible
    $\sigma_{m'}\in\Hom(\one,m').$ Using the same trick, we see that
    $\sigma_m\otimes \sigma_{m'}\in\Hom(\one,mm')$ is the direct sum of two
    invertible 1-morphisms. Thus, $mm'=\one$. Together with $m^2=\one$ we
    conclude that $m=m'$.
  \item $m$ is central, $ \forall g, mg=gm$. To see this,
    consider $\one_g\otimes \sigma_m\otimes\one_{g^{-1}} $ which is a non-invertible
    1-morphism in $\Hom(gg^{-1},gmg^{-1})=\Hom(\one,gmg^{-1})$. Since $m$ is
    unique we must have $m=gmg^{-1}$.
\end{enumerate}
Therefore, it is possible to have a $Z_2$ string $m$ which can be open on the
canonical boundary of EF topological orders. Its end points (non-invertible
1-morphism in $\Hom(\one,m)$) have quantum dimension $\sqrt{2}$. 

Physically, $m$ string is distinguished from the trivial string under the
equivalences generated by local unitary transformations.  In other words $m$
string and trivial string have different l-types.  $m$ string becomes the same
as the trivial string under the equivalences generated by non-local unitary
transformations.  So $m$ string and trivial string have the same type.
This implies that $m$ is a descendant string formed
by lower dimensional topological excitations (since it can have boundary).  
On the boundary of a EF topological order, the only lower
dimensional topological excitations are the trivial particles and the fermions.
Since there is no topological order in 1D, the trivial particles cannot form
any non-trivial strings.  On the other hand, the fermions can form topological
$p$-wave superconducting chain,\cite{K0131} called the Majorana chain.  Thus
the $m$ string must be a  Majorana chain.  The 1-morphism between $m$ string
and trivial string in $\Hom(\one,m)$ (\ie the end point of $m$ string) is the
Majorana zero mode at the end of the Majorana chain.

We would like to emphasize here that such extra string $m$ and non-invertible
1-morphism $\sigma_m$ are the only remaining possibility beyond the case
discussed in the last section.  The boundary strings are labeled by a larger
group $\hat G_b$, which is a central $Z_2$ extension of $G_b$,
\[\{\one,m\}\equiv Z_2^m \to \hat G_b \stackrel{\pi^m}{\to} G_b.\] With the
enlarged boundary string types and non-invertible 1-morphism, EF
topological orders are classified by unitary fusion 2-categories $\sA_b^3$
described in Section \ref{bndry}.

\subsection{Properties of the unitary fusion 2-categories}

Next we discuss in more detail how the extra string $m$ and non-invertible
1-morphism $\sigma_m$ will affect the classification results.  

Now, strings are labeled by a larger group $\hat G_b$ on the canonical
boundary. But note the fact that the data and conditions not involving
$\sigma_m$ are not affected at all. This means that we can start with a
solution $(\hat G_b, \hat e_2, \hat n_3, \hat {\nu}_4)$ to \eqref{o4gb} with
the larger group, and then deal with the additional constraints involving
$\sigma_m$.

The $\sigma_m$ 1-morphism must itself satisfy some additional braiding and
fusion constraints. This means that $\hat {\tilde
b}(\bullet,\bullet,\bullet,\bullet)$ and $\hat {\tilde
n}_3(\bullet,\bullet,\bullet)$ involving $\sigma_m$ take different forms. We
expect that the results are closely related to the braiding statistics of Ising
anyons.

Besides, the strings of l-types $g$ and $gm$ can be ``connected'' by
non-invertible 1-morphisms. This implies, for example, that $\hat {n}_3(g,h,j)$
and $\hat n_3(gm,h,j)$, or $\hat {\nu}_4(g,h,j,k)$ and $\hat {\nu}_4(g,
hm,jm,k)$, etc., are related by $m$ and $\sigma_m$. As a result, $\hat n_3$ and
$\hat {\nu}_4$ can be factorised, $\hat n_3= n_3 + n_m, \hat {\nu}_4={\nu}_4
\nu_m$ where $ n_3, \nu_4$ are cochains in $G_b=\hat G_b/Z_2^m$, and
$n_m,\nu_m$ are factors depending on how the $m$ string is attached.


In other words, there is map from the unitary fusion 2-categories
$\sA_b^3$ that classify EF topological orders to the pointed unitary fusion
2-categories $\bar\sA_b^3$ that classify EF1 topological orders.  Such a
map sends a unitary fusion 2-category $\sA_n^3$ with objects $\hat
G_b$ to a pointed unitary fusion 2-category $\bar\sA_n^3$ with objects $\hat
G_b$, by taking the pointed sub-2-category (ignoring the non-invertible
1-morphisms).  Therefore, there is map from EF topological orders to
EF1 topological orders, which sends a EF topological order with pointlike
excitations described by $\sRep(Z_2^f\gext G_b)$ to a EF1 topological order with
pointlike excitations described by $\sRep(Z_2^f\gext \hat G_b)$.  This
relation allows us to obtain a EF topological order with pointlike
excitations $\sRep(Z_2^f\gext G_b)$ from a EF1 topological order
with pointlike excitations $\sRep(Z_2^f\gext \hat G_b)$ that satisfies
certain additional constraints.

We leave the details of the additional constraints involving the
non-invertible 1-morphism $\sigma_m$
for future work (see \Ref{ZLW}). We believe that they are the same as those for
fermionic SPTs with the Majorana chain layer.

\subsection{Majorana zero modes at triple-string intersections}

\begin{figure}[tb] 
\centering \includegraphics[scale=0.8]{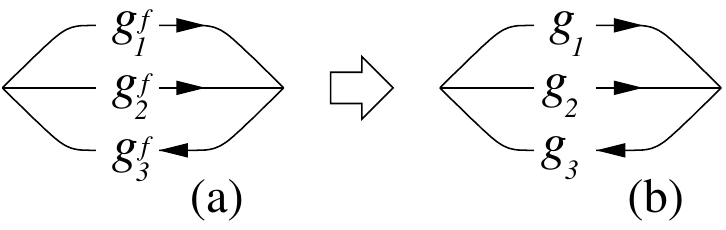} 
\caption{
(a) A string configuration in bulk, described by the conjugacy class of a
triple $(g_1^f,g_2^f,g_3^f)$ in $G_f$.  (b) Moving to the boundary, the string
configuration turns into one is labeled be three group elements $(g_1,g_2,g_3)$
in $\hat G_b$.
}
\label{3strings1} 
\end{figure}

In the following, we
will describe a bulk property that allow us to distinguish the  EF1 and EF2
topological orders.  In particular we will design a setup which allows us to
use the appearance of Majorana zero mode to directly measure the cohomology
class of $\rho_2$.  For simplicity, let us assume $G_f$ to Abelian for the time
being. In this case, the different types of bulk strings are labeled by $g^f
\in G_f$.  In our setup, we first choose a \emph{fixed set} of trapping
potentials that trap a fixed set of strings labeled by $g^f \in G_f$.  Note
that the different strings in the set can all be distinguished by their
different braiding properties with the pointlike excitations.  Then, choosing
three strings from such a fixed set, we can form a configuration in Fig.
\ref{3strings1}a.  For Abelian $G_f$, one may expect that the degeneracy for
the configuration Fig.  \ref{3strings1}a to be 1.  In the following, we will
show that, sometimes the configuration Fig.  \ref{3strings1}a has a
2-fold topological degeneracy.
By measuring which triples $g_1^f,g_2^f,g_3^f$ in the fixed set of strings give
rise to 2-fold topological degeneracy, we can determine the cohomology class of
$\rho_2$ directly.

One may point out that the appearance of 2-fold topological degeneracy is not
surprising at all, since the EF topological order with Abelian $G_f$ contains
an emergent fermion in the bulk that has an unit quantum dimension. Such
fermions can form a Majorana chain.\cite{K0131}  Some strings in the fixed set
may accidentally carry such a Majorana chain.  If one or three strings in the
configuration Fig.  \ref{3strings1}a carry Majorana chain, then the
configuration will have a 2-fold topological degeneracy, coming from the two
Majorana zero modes at the two intersection points.  So it appears that the
appearance of 2-fold topological degeneracy in the configurations Fig.
\ref{3strings1}a is not a universal property.  We can remove the 2-fold
topological degeneracy by choosing our fixed set of strings properly such that
none of the string in the fixed set carry Majorana chain. This indeed can be
achieved when $\rho_2$ is a coboundary.  When $\rho_2$ is a non-trivial
cocycle, there is an obstruction in determining if a string carries a Majorana
chain or not.  As a result, no matter how we choose the fixed set of strings,
there are always some triples $g_1^f,g_2^f,g_3^f$ in the fixed set of strings,
such that their configurations Fig.  \ref{3strings1}a have 2-fold topological
degeneracies.

How to determine $\rho_2$ from the topological degeneracy of the configurations
Fig. \ref{3strings1}a?  We first measure the topological degeneracy Fig.
\ref{3strings1}a where the three strings are chosen from the fixed set.  If
there is a 2-fold topological degeneracy, we assign
\begin{align}
\rho_2^f(g_1^f,g_2^f) =-1.
\end{align}
If there is no degeneracy,
we assign
\begin{align}
\rho_2^f(g_1^f,g_2^f) =1.
\end{align}
From the function $\rho_2^f(g_1^f,g_2^f)$ 
we can determine the cohomology class of $\rho_2 \in H^2(G_b,Z_2^m)$.

To see this, we first move the string  configuration to the boundary.  In this
case, the bulk string labeled by $G_f$ first have a reduction from $G_f
\stackrel{\pi^f}{\to} G_b$, and then an extension to $\hat G_b$.  In other
words, the bulk string types $g^f_1$, $g^f_2$, and $g^f_3$ in $G^f$ change to
the boundary string types $g_1$, $g_2$, and $g_3$ in $\hat G_b$ (see Fig.
\ref{3strings1}b), which satisfy
\begin{align}
 \pi^f(g^f_i)=\pi^m(g_i) \in G_b,
\end{align}
where $\pi^f$ and $\pi^m$ are the projections $G_f \stackrel{\pi^f}{\to} G_b$
and $\hat G_b \stackrel{\pi^m}{\to} G_b$.

We note that the elements in $\hat G_b$ can be labeled as $(g^b,x)$, $g^b \in
G_b$ and $x \in Z_2^m$.  The multiplication in $\hat G_b$ is given by
\begin{align} 
(g^b,x) (h^b,y)=(g^bh^b, \rho_2(g^b,h^b) xy) 
\end{align} 
where $\rho_2(g^b,h^b)$ is a group 2-cocylce in $H^2(G_b,Z_2^m)$.  Thus $g_i$
has a form $(g^b_i,x_i)$ where $g_i^b=\pi^f(g^f_i)$.  Here we like to stress
that the bulk string $g^f_i$ only determines the $g^b_i$ component in the pair
$(g^b_i,x_i)$.  Since we move the fixed set of bulk strings to the boundary in
a particular way, we obtain a particular $x_i$ for each $g^b_i$.
In other words, $x_i$ is a function of $g^b_i$, denoted by
\begin{align}
 x_i=x(g^b_i).
\end{align}

Although the bulk string types satisfy $g^f_1g^f_2 = g^f_3$ which leads to
$g^b_1g^b_2 = g^b_3$,  the boundary string types $g_i$, as a particular
lifting from $G_b$ to $\hat G_b$ may not satisfy $g_1 g_2 = g_3 $.  In fact we
have 
\begin{align}
 [g^b_1, x(g^b_1)] [g^b_2, x(g^b_2)] &= 
[g^b_1g^b_2, \rho_2(g^b_1,g^b_2) x(g^b_1)x(g^b_2)]
\nonumber\\
&
 =[g^b_3, \t\rho_2(g^b_1,g^b_2) x(g^b_3)]
\end{align}
where 
\begin{align}
 \t\rho_2(g^b_1,g^b_2) = \rho_2(g^b_1,g^b_2) x(g^b_1)x(g^b_2)x^{-1}(g^b_1
 g^b_2) .\label{trho2}
\end{align}
When $\t\rho_2(\pi^f(g^f_1), \pi^f(g^f_2))=m$, we have
$g_1g_2=mg_3$ and the intersection point
will carry a Majorana zero mode.  In other words, the boundary configuration
Fig. \ref{3strings1}b has a 2-fold topological degeneracy if
$\t\rho_2(\pi^f(g^f_1), \pi^f(g^f_2))=m$.

Since the boundary configuration Fig. \ref{3strings1}b can be a short distance
away from the boundary, thus moving to the boundary represents a weak
perturbantion. In this case, the boundary configuration Fig. \ref{3strings1}b
having a 2-fold degeneracy implies that the corresponding bulk configuration
Fig. \ref{3strings1}a also has a 2-fold degeneracy.  
In other words
\begin{align}
  \t\rho_2(\pi^f(g^f_1), \pi^f(g^f_2)) = \rho_2^f(g^f_1, g^f_2).\label{rhoftrho}
\end{align}
We see that the cocycle $\t\rho_2$ can be determined by measuring the
topological degeneracy for bulk string configurations Fig. \ref{3strings1}a.
We note that $\t\rho_2$ and $\rho_2$ differ by a coboundary \eqref{trho2}. Thus, up to a
coboundary, $\rho_2$ can be determined by measuring the topological degeneracy
for bulk string configurations Fig. \ref{3strings1}a.

We like to pointed out that even when $G_f$ is non-Abelian, a non-trivial
$Z_2^m$ extension $\rho_2$ also gives rise the Majorana zero modes for some
triple string intersections.  But in this case, there are extra topological
degenercies on intersections of three strings coming from the non-Abelianness
of $G_f$.  The  appearance of topological degenerates does not directly imply
the appearance of Majorana zero modes.  It is non-trivial to separate which
topological degeneracy comes from non-Abelian $G_f$ and which comes from
Majorana zero modes.  
However, the similar results also hold for non-Abelian $G_f$.  In the
following, we will describe those results for non-Abelian $G_f$, but now from a
pure bulk point of view.
  
Again, the key step is to choose a \emph{fixed set} of trapping potentials that
trap a fixed set of strings labeled by $\chi_{g^f} \subset G_f$.  Here $\chi_{g^f}$ is
the conjugacy class that contains $g^f \in G_f$.  We stress that the different
strings in the set can all be distinguished by their different braiding
properties with the pointlike excitations.  We call two strings to be
equivalent if they have the same brading properties with all the pointlike
excitations.  Thus the strings in our fixed set are all inequivalent.  We also
assume our fixed set is complete, in the sense that it contains all
inequivalent strings.  In other words, the number of strings in the set is
equal to the number of conjugacy classes in $G_f$.

We note that condensation of the pointlike excitation can also form a
stringlike excitation. For example condensation of $Z_2$-charges along a chain
in a $Z_2$ gauge theory can form a stringlike excitation that have trivial
braiding with all the pointlike excitations.  We call such kind of stringlike
excitations descendant stringlike excitations, which all equivalent to trivial
string under non-local unitray transformations on the string.  
The above $Z_2$-charge condensed chain has a 2-fold degeneracy since
it is like a $Z_2$ symmetry breaking state.  As a result, the corresponding
descendant stringlike excitation has a quantum dimension 2 (and such a
quantum-dimension-2 string is equivalent to a trivial string with quantum
dimension 1).  We point out that our fixed set of strings do not contain
strings that only differ by attaching a descendant stringlike excitation, since
they are regarded as equivalent. 

But each string in the fixed set may carry some additional descendant stringlike
excitations.  We like to reduce this ambiguity by requiring the strings in the
fixed set do not carry descendant strings.  This is achieved by replacing each
string in the set by its equivalent string that have a minimal quantum
dimension.  However, this still does not remove all the ambiguity.

When and only when $G_f$ has a form $G_f=Z_2^f \times G_b$, the following two
facts become true: (1) there are bulk fermionic excitations with unit quantum
dimension, and (2) the condensation of such fermions only break the $Z_2^f$
symmetry \cite{KW1485} but not any other symmetries in $G_b$.  Such fermion
condensed chain is nothing but the Majorana chain.\cite{K0131} The Majorana
chain is a descendant string.  But amazingly, despite the $Z_2^f$ symmetry
breaking on open Majorana chain, the closed Majorana chain has no ground state
degeneracy and the Majorana chain has a quantum dimension 1.  Attaching
Majorana chain to a string will not change the quantum dimension of the string.
So the strings in our fixed set, even after minimizing the quantum dimensions,
may still carry Majorana chains.  It turns out that there is an obstruction to
find a complete set of inequivalent strings that do not carry Majorana chains
for EF2 topological orders, while for EF1 topological orders there is no such
an obstruction.

To test if the strings in our fixed set carry Majorana chains or not, we
choose three strings from our fixed set to form a configuration in Fig.
\ref{3strings}.  The topological degeneracy of the configuration is calculated
in the following way.  We first consider a set of pairs that have a form $(\t
g_1,\t g_2)$, where $\t g_1 \in \chi_{g^f_1}$ and $\t g_2 \in \chi_{g^f_2}$.
The two pairs
$(\t g_1,\t g_2)$ and
$(\t g_1',\t g_2')$ are equivalent if they are related by
\begin{align}
 \t g_1' = h \t g_1 h^{-1},\ \ \ \
 \t g_2' = h \t g_2 h^{-1},\ \ \ \ h\in G_f.
\end{align}
The number of equivalent classes of the pairs, $N(\chi_{g^f_1},\chi_{g^f_2})$, is the
topological degeneracy of the configuration in Fig.  \ref{3strings}, provided
that the three strings do not carry  Majorana chains.  If one or three strings
carry  Majorana chains, the topological degenercy of the configuration in Fig.
\ref{3strings} will be given by $2N(\chi_{g^f_1},\chi_{g^f_2})$.  In this case, we say
the triple string intersection in Fig.  \ref{3strings} carry a Majorana zero
mode.

Now we introduce a function: $\rho_2^f(g_1^f,g_2^f)=1$
if the topological degeneracy of the configuration in Fig.  \ref{3strings} is
$N(\chi_{g^f_1},\chi_{g^f_2})$, and $\rho_2^f(g_1^f,g_2^f)=-1$ if the topological
degeneracy is $2N(\chi_{g^f_1},\chi_{g^f_2})$.  Clearly $\rho^f_2$ satisfies
\begin{align}
\label{rhofinv}
 \rho^f_2(g_1^f,g_2^f) = \rho^f_2(h_1 g_1^f h_1^{-1},h_2 g_2^f h_2^{-1}),\ \
h_1,h_2 \in G_f.
\end{align}
$\rho_2^f$ in the above is a cocycle in $Z^2(G_f,Z_2^m)$.  If $\rho_2^f$
is a coboundary, we can choose a fixed set of strings such that all the
triple string intersections do not carry Majorana zero modes.  The corresponding
bulk topological order is an EF1 topological order.  If $\rho_2^f$ is a
non-trivial cocycle, then for any choice of a fixed set of strings, there are
always triple string intersections that carry Majorana zero modes.  The
correspond bulk topological order is an EF2 topological order.

\begin{figure}[t]
  \centering
  \includegraphics[scale=0.6]{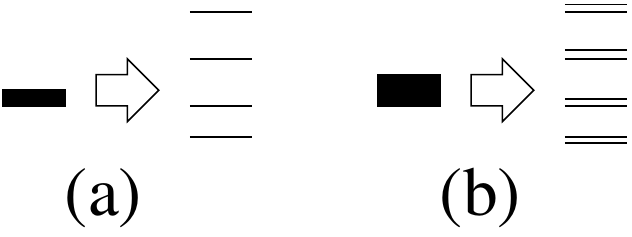}
  \caption{
The splitting of the topological degeneracy as we move
 string configuration Fig. \ref{3strings} to wards the canonical boundary.
(a) the case for topological degeneracy $N(\chi_{g^f_1},\chi_{g^f_2})$.
(b) the case for topological degeneracy $2N(\chi_{g^f_1},\chi_{g^f_2})$.
}
  \label{Nsplit}
\end{figure}

The existence of the canonical boundary for a EF topological order requires 
$\rho^f_2(g_1^f,g_2^f)$ to be a function on $G_b$, \ie it has a form
\begin{align}
\label{rhofrho}
 \rho^f_2(g_1^f,g_2^f) = \t \rho_2[\pi^f(g_1^f),\pi^f(g_2^f)] ,
\end{align}
where $\t \rho_2 \in Z^2(G_b,Z_2^m)$.  To understand the above result, we move
the string configuration Fig.  \ref{3strings} towards the canonical boundary.
The string type will change from the bulk type $\chi_{g^f}$ to the boundary l-type
$g\in \hat G_b$: $\chi_{g^f} \to g^b \to g$ that satisfy
\begin{align}
 g_b \in \pi^f(\chi_{g^f}),\ \ \ \ \ g_b = \pi^m(g).
\end{align}
The $N(\chi_{g^f_1},\chi_{g^f_2})$-fold or $2N(\chi_{g^f_1},\chi_{g^f_2})$-fold
topological degeneracy will split (see Fig. \ref{Nsplit}).  Note that the
2-fold topological degeneracy from Majorana zero modes is not affected by
moving to the boundary.  Because of the reduction $\chi_{g^f} \to g^b$ on the
boundary, the Majorana zero modes can only depend on $G_b$, and hence
$\rho^f_2(g_1^f,g_2^f)$ is only a function on $G_b$.  The resulting
$\t\rho_2(g^b_1,g_2^b)$ determines the $Z_2^m$ extension of $G_b$.



\subsection{Necessary conditions for EF2 topological order}

%
From the bulk consideration in the last section, we see that the $\rho_2$ 
characterizing the EF2 topological orders are highly restricted. We focus on
the particular $\tilde \rho_2$ that directly comes from measuring the
Majorana zero modes in the bulk; it can differ from $\rho_2$ by a coboundary.  First, the
pullback of $\tilde \rho_2$ by $G_f \xrightarrow{\pi^f} G_b$ gives us a
$\rho^f_2=(\pi^f)^* \tilde \rho_2 \in H^2(G_f,\Z_2)$ (see \eqn{rhoftrho}).  Such a
pullback must satisfy \eqn{rhofinv}. This gives us a condition on $\t\rho_2$:
\begin{align}
\tilde\rho_2(g_1^b,g_2^b) = \tilde\rho_2(h_1 g_1^b h_1^{-1},h_2 g_2^b h_2^{-1}),\ \ \ \
h_1,h_2 \in G_b.\label{rho2c}
\end{align}
In other words, EF2 topological order can exist only when $G_b$ has non-trivial
2-cocycles with the above symmetry condition.  This is the first necessary
conditions for EF2 topological orders.  We note that when $G_b$ is abelian, the
above condition becomes trivial and imposes no constraint.

We also like to point out that a Majorana chain can be attached to a bulk
string characterized by the conjugacy class $\chi_g$ of $G_f$ only when the
centralizer group $Z_g(G_f)$ is a trivial $Z_2^f$ extension.
Here $Z_g(G_f)$ is the subgroup that commutes with an element $g$
in the conjugacy class $\chi_g$
\begin{align}
  Z_g(G_f) =\{ x\in G_f| gx=xg\}.
\end{align}
Physically, the bulk string $\chi_g$ breaks the ``symmetry'' of the particles
from $G_f$ down to $Z_g(G_f)$.  If $Z_g(G_f)$ is not a trivial $Z_2^f$
extension, then a fermion condensation that breaks the $Z_2^f$ ``symmetry''
must also break some additional ``symmetries''.  In this case, we cannot
attach  Majorana chain to the bulk string $\chi_g$, since the Majorana chain
corresponds to a fermion condensation that breaks only the $Z_2^f$
``symmetry''.\cite{KW1485}  

Let us introduce a M-function on $G_f$
\begin{align}
 M(g)=
\begin{cases}
 0,  \text{ $Z_g(G_f)$ is a trivial $Z_2^f$ extension} \\
 1,  \text{ otherwise} \\
\end{cases}
\end{align}
Since 
\begin{align}
 Z_g(G_f) = Z_{zg} (G_f),
\end{align}
where $z$ is the generator of $Z_2^f$, we have
\begin{align}
 M(g)=M(zg).
\end{align}
Therefore, we may also view $M$ as a function on $G_b$.

Since the bulk string $\chi_g$, $g\in G_f$,
has no ambiguity of Majorana string when $M(g)=1$,
we see that $\rho^f_2$ satisfies
\begin{align}
 \rho^f_2(g_1^f,g_2^f)=0, \text{ if }
M(g_1^f)=M(g_2^f)=M(g_1^fg_2^f)=1.
\end{align}
This becomes a condition on the $G_b$-cocycle $\tilde\rho_2$
\begin{align}
 \tilde\rho_2(g_1^b,g_2^b)=0, \text{ if }
M(g_1^b)=M(g_2^b)=M(g_1^bg_2^b)=1.
\label{rho2M}
\end{align}

This is the second necessary conditions for EF2 topological orders.
We note that the two conditions \eqref{rho2c}\eqref{rho2M} are not invariant
under adding coboundaries. Physically, on the canonical boundary, unlike in
the bulk, it is always possible to attach Majorana chains to strings, since the
$G_f$ ``symmetry'' is broken down to $Z_2^f$ on the boundary.
This can change $\rho_2$ by arbitrary coboundaries. Thus, generic $\rho_2$ may not
satisfy \eqref{rho2c}\eqref{rho2M}; we only require \eqref{rho2c}\eqref{rho2M}
for a particular $\tilde \rho_2$ that is cohomologically equivalent to generic
$\rho_2$.

As an example, for $G_f=Z_4^f\times G_b'$, we find $M(g)=1$ for all $g\in
Z_4^f\times G_b'$.  Thus, there is no EF2 topological order with
$G_f=Z_4^f\times G_b'$.  In \Ref{LZW}, it was shown that 3+1D fermionic
$Z_4^f$-SPT phases from fermion decoration are described by $\Z_2$.  The above
argument shows that there is no Majorana chain decoration for $Z_4^f$ symmetry.
Thus fermion decoration produces all SPT phases, and all 3+1D fermionic
$Z_4^f$-SPT phases are classified by $\Z_2$.


\section{A general framework for 3+1D topological orders with symmetries}

We see that in 3+1D the intrinsic topological orders are closely related to SPT
phases. In the above section we showed that the classification of EF
topological orders is the same as that of fermionic SPT phases. Without the
Majorana chain layer, both EF topological orders and fermionic SPT phases are
classified by the group super-cohomology theory; with the Majorana chain layer,
also very strong evidence indicates that they have one-to-one correspondence.
Combined with our previous results on 3+1D AB topological orders,
we conclude that \myfrm{All 3+1D topological orders correspond to gauged 3+1D
SPT phases: AB topological orders correspond to gauged bosonic SPTs and EF
topological orders correspond to gauged fermionic SPTs.}

The SPT and the topological order are the end points of ungauging/gauging
procedures respectively. They are also the two extreme cases with only symmetry
no intrinsic topological order and only intrinsic topological order no
symmetry.  Because of these, it is natural to conjecture that if we partially
gauge a SPT or ungauge a topological order, in-between we should get a state
with both symmetry and topological order, in other words, a symmetry enriched
topological order (SET). Therefore, we expect the following general
classification framework for 3+1D topological phases with symmetries:
\begin{align*}
  \xymatrix{&*+[F]{\text{SPT}}\ar[rd]^{\text{gauging}}
  \ar[ld]_{\text{gauging}}&\\
  *+[F]{\text{SETs}}\ar[rd]_(.4){\text{gauging}}
  &\cdots&*+[F]{\text{SETs}}\ar[ld]^(.4){\text{gauging}}\\
  &*+[F]{\text{Topological order}}&
  }
\end{align*}
Different partially gauging procedures, equivalently different subgroup
sequences $H_1\subset H_2\subset \cdots\subset G$, give rise to different
sequences of intermediate SETs. The starting point, SPT, and end point,
topological order, are fixed. They have one-to-one correspondence between each
other, according to our classification results. We believe that in the same
gauging sequence the phases share the same classification data. However, their
physical interpretations are different at different steps.

In particular, fermionic SETs and topological orders (note that EF topological
order is a \emph{bosonic} topological order with emergent fermionic particles)
should be special cases starting from fermionic SPTs but keep the fermion
number parity (FNP) not gauged until the last step:
\begin{align*}
  \xymatrix{
  *+[F]{\text{fermionic SPT}}\ar[d]^{\text{gauging (keep FNP)}}\\
  *+[F]{\text{fermionic SETs}}\ar[d]^{\text{gauging (keep FNP)}}\\
  *+[F]{\text{fermionic topological order}}\ar[d]^{\text{gauging FNP}}\\
  *+[F]{\text{EF topological order}}
}
\end{align*}

Recall that in 2+1D we classified topological phases with symmetry by a triple
of categories $\sE\subset\sC\subset\sM$\cite{LKW1602.05936,LKW1602.05946} where $\sE$ is the symmetric category
of local excitations and corresponds to the representations of the symmetry
group, $\sE=\Rep(G)$ or $\sE=\sRep(G_f)$, $\sC$ is the category of all bulk
excitation and $\sM$ is the gauged theory. In particular for 2+1D SPT phases we
have $\sE=\sC\subset\sM$.
Now this idea naturally generalizes to 3+1D, since any 3+1D topological order
contains a symmetric subcategory $\sE$ corresponding to its pointlike
excitations, and can be viewed as a gauged SPT $\sM$ with symmetry $\sE$. A
generic 3+1D SET is then described by certain 2-category $\sC$ satisfying
$\sE\subset\sC\subset\sM$. In the gauging procedures, the modular extension
$\sM$ remains the same, while $\sE$ and $\sC$ becomes smaller and larger
respectively ($\sE=\sC=\Rep(G)$ or $\sRep(G_f)$ for the SPT phase while $\sE$
is trivial and $\sC=\sM$ for the topological order).

As we already have good understanding about the 3+1D SPT phases, it is thus
quite hopeful for a complete understanding of 3+1D topological orders and
symmetries by thoroughly studying the (partially) gauging procedures.

%
%
%
%
%

\bigskip

We thank Zheng-Cheng Gu for helpful discussions, and Chenjie Wang for sharing
his unpublished result about the appearance of Majorana zero modes on linked
loops in some 3+1D fermionic SPT states.  XGW is supported by NSF Grant No.
DMR-1506475 and DMS-1664412. Part of the work was done at Perimeter Institute
for Theoretical Physics and Tsinghua University. Research at Perimeter Institute is supported by
the Government of Canada through Industry Canada and by the Province of Ontario
through the Ministry of Research. TL's visit at Tsinghua University was
supported by Chinese Ministry of Education under grants No.20173080024.

\appendix

\section{Tannaka Duality}

\label{TD}

Our approach in this paper relies heavily on the Tannaka duality\cite{T3801},
or Tannaka reconstruction theorem for group representations.  In this section,
we will give a physical introduction of Tannaka duality. In the meantime, we
will also introduce and explain some important concepts used in this paper in
detail.

\subsection{Two physical models}

A physical motivation of the Tannaka Duality is the following: let us consider
a bosonic or a fermionic system with a symmetry $G$. We assume the ground state
to be a product state that does not break the symmetry. If we only measure the
system via probes that do not break the symmetry, can we detect the symmetry
group of the system?  We note that a symmetry transformation acts on objects
that break the symmetry (\ie not invariant under the symmetry transformation).
Thus we need to break the symmetry in order to measure the symmetry
transformation directly.  In contrast, the symmetric probes only produce
objects that do not break the symmetry, such as particles trapped by symmetric
potential that are described by representations $\rho$ of the symmetry group:
$\rho\in $ Rep$(G)$. On the other hand, the symmetric probes do allow us to
fuse and braid those symmetric particles in arbitrary ways.

To describe those fusion and braiding processes, the concept of fusion space is
important: if the particles are obtained by symmetric trap potentials, then the
fusion space $\cV$ is simply the ground state subspace of the total Hamiltonian
with traps: $H_\text{tot}=H_0+\sum_i \Del H_\text{trap} (x_i)$ which trap
particles $p_i$ at $x_i$. We denote  the fusion space as
$\cV(M,p_1,p_2,\cdots)$ where $M$ is the space manifold that supports our
system.  So the fusion and the braiding processes, as well as the symmetric
deformation of the Hamiltonians $H_0$ and $\Del H_\text{trap}$, correspond to
unitary linear maps on the fusion space.  Tannaka duality tells us how to use
those symmetric operations, \ie the linear maps on the fusion space
$\cV(M,p_1,p_2,\cdots)$, to obtain the symmetry group $G$.

Mathematically, the fusion and braiding, as well as the symmetric deformation
of the Hamiltonians $H_0$ and $\Del H_\text{trap}$, on all the possible trapped
particles form a structure which is denoted as $\Rep(G)$ if the all the
particles are bosons, or  as $\sRep(G)$ if the some particles are fermions.
Such a structure is called symmetric fusion category (SFC).  The particles are
labeled by the representations of $G$, which form a set Rep$(G)$.  So a SFC
$\Rep(G)$ or $\sRep(G)$ contains the set Rep$(G)$ whose elements are called
objects (which correspond to trapped particles). $\Rep(G)$ or $\sRep(G)$ also
contains addition data that describe fusion and braiding of particles in
Rep$(G)$.  In particular, the fusion of the particles are non-trivial, since
the particles are described by the representations of $G$, and the fusion of
the representations is non-trivial.

If we just know the set of representations Rep$(G)$, we cannot recover the
group $G$. But if we also know all symmetric operations, such as fusion and
braiding, as well as the symmetric deformation of the Hamiltonians $H_0$ and
$\Del H_\text{trap}$; in other words, if we know $\Rep(G)$ or $\sRep(G)$, then
according to Tannaka duality, we can recover the group $G$.

Another physical motivation of the Tannaka Duality is more relevant to this
paper.  We consider a 3+1D topological order $\sC^4$.  The pointlike
excitations in the  topological order are bosons or fermions with trivial
mutual statistics.  Those particles have a non-trivial fusion rule.  The fusion
and braiding of those particles are also described by a SFC $\sE$.  Tannaka
duality tells us that from $\sE$,  we can recover a group $G$.  Thus each 3+1D
topological order contains a hidden group $G$.  In this second example, we do
not even have a symmetry. All the operations, such as fusion, braiding, and
deformation of $H_0$ and $\Del H_\text{trap}$, are allowed, as long as they are
generated by local interaction.  But how can one recover a group from a problem
that has no symmetry?

In the first example, we do have  symmetry, but we want to recover the symmetry
group via the symmetric operations. In the second example, we want to recover
the hidden group in 3+1D topological order which has no symmetry. This two
problems happen to be the same problem, which is solved by Tannaka duality.


\subsection{Tannaka duality I: all boson} 

\subsubsection{Statement of Tannaka duality}

For the moment we restrict to an all-boson SFC $\sE$.  Mathematically, Tannaka
duality states that we can reconstruct a group $G$ from SFC $\sE$ by the
automorphisms of a fiber functor, namely a braided monoidal functor $F$, from
$\sE$ to the category of vector spaces, $\Ve$
\begin{align}
&  G\equiv\Aut(F:\sE \to\Ve)
,
\end{align}
and Tannaka duality tells us that
\begin{align}
  \sE\cong \Rep(G).
\end{align}
This is how we find the hidden group in a SFC $\sE$.

To understand Tannaka duality let us choose the SFC to be the category formed
by the representations of a finite group $\Rep(G)$.  We like to find out
what is the automorphisms of a fiber functor $\t G\equiv\Aut(F:\sE \to\Ve)$?

Let us first describe the representation category $\Rep(G)$:
\begin{enumerate}
\item
An object in $\Rep(G)$ is a group representations $p$, which corresponds to a pair $p\equiv (V(p),\rho_p)$, where $V(p)$ is a vector space
equipped with a group action $\rho_p:G\to\mathrm{GL}[V(p)]$.  The set of
objects in $\Rep(G)$ is formed by all such pairs (\ie by all the group
representations).  
\item
The morphisms in the SFC $\Rep(G)$, $p'\to p$, correspond to
the embedding map $u: V(p') \to V(p)$ which commutes with the group action,
$\rho_{p}(g) u = u \rho_{p'}(g)$.
The morphisms allow us to define the notion
of simple objects which correspond to irreducible representations.
\item
Representations can be ``fused'' $p_1\otimes p_2$, which corresponds to taking
the tensor product of the vector spaces $V({p_1})\otimes_\C V({p_2})$ and the
new group action is $\rho_{p_1\otimes
p_2}(g)=\rho_{p_1}(g)\otimes_\C\rho_{p_2}(g)$:
\begin{align}
p_1\otimes p_2 = (V({p_1})\otimes_\C V({p_2}),\rho_{p_1}(g)\otimes_\C\rho_{p_2}(g)). 
\end{align}  
\end{enumerate}

In this case, we have the forgetful functor that maps a representation category
$\Rep(G)$ to the category of vector spaces $\Ve$,
$F:p\equiv(V(p),\rho_p)\mapsto V(p)$ (forgetting the group action part), which
is called a fiber functor.  An automorphisms of such a fiber functor $F$ is a
set of unitary maps, $\al =\{\al_p\}$, one map for each $p$ and $\al_p$ acts on
$V(p)$.  Such set of maps must be compatible with the fusion rule described
above, as well as the morphisms $p'\to p: V(p') \overset{u}{\to} V(p)$, \ie
satisfying $ \al_{p} u = u \al_{p'}$.  The set of all those automorphisms form
a group
\begin{align}
 \al\cdot \al'=\{\al_p\}\cdot \{\al_p'\}=\{\al_p\al_p'\}.
\end{align}
Such a group is the automorphism group, which happen to be $G$:
\begin{align}
  G\cong \Aut(F:\Rep(G)\to \Ve).
\end{align}
This is because, to be compatible with the morphisms and the fusion rule,
$\al_p$ has to be $\rho_p(h)$ for a certain $h\in G$.  In fact, this is how we
recover the symmetry group $G$ in the first model.

In the following, we will describe Tannaka's construction and the above
statements, in terms of the two physical models described above, where the
particles are described by a SFC $\sE$.  This way, one may gain a more physical
understanding of Tannaka duality.

\subsubsection{ Irreducible representations from symmetry operations}

Before trying to obtain the group, let us try to obtain the irreducible
representations of the group first.  In general, a particle $p \in \sE$
(trapped by a symmetric potential in the first model) corresponds to a
representation. But which particles correspond to irreducible representations?
To address this question, we start with the fusion space of $p$ with other
particles $\cV(M,p,q,\cdots)$.  Note that $\cV(M,p,q,\cdots)$ is the ground
state subspace of $H_0+\Del H_\text{trap}(x_p) +\Del H_\text{trap}(x_q)
+\cdots$ that traps the particle $p$ at $x_p$, particle $q$ at $x_q$, \etc.  By
deforming (or deforming while preserving the symmetry for the first model) just
$\Del H_\text{trap}(x_p)$ to $\Del H_\text{trap}'(x_p)$, we may split the
ground state degeneracy
\begin{align}
 \cV(M,p,q,\cdots) =\cV_1 \oplus\cV_2\oplus \cdots.
\end{align}
the new ground state subspace $\cV_1$ can be viewed as the fusion space of
another particle $p'$ at $x_p$ with other particles $q,\cdots$, $\cV_1 =
\cV(M,p',q,\cdots)$. Thus the above splitting of $ \cV(M,p,q,\cdots)$ can be
rewritten as
\begin{align}
\label{cVXXp}
 \cV(M,p,q,\cdots) = \cV(M,p',q,\cdots)\oplus \cV_2\oplus \cdots.
\end{align}
Then we say that there is a morphism from $p'$ to $p$: $p'\to p$.\footnote{In
  mathematical formulation of fusion categories, this $p'\to p$ morphism is
  only an embedding morphism; generic
  morphisms are generated by linear combinations, compositions, taking Hermitian
  conjugates of these
  embeddings. But in physics we always (secretly) assume linearity and
  unitarity (including taking Hermitian conjugates of morphisms). So taking
only the embedding as our (basis) morphism suffices for our purpose.}   Here, a
morphism $p'\to p$ can be understood as that
the fusion space of $p'$, after a proper unitary transformation, is contained in the
fusion space of $p$.  If we have morphisms in both directions $p'\to p$ and
$p \to p'$, then the fusion space of $p$ is the same as the fusion space of
$p'$, up to an unitary transformation.  If $p'\to p$ implies $p\to p'$,
for all $p'$'s, then the fusion space of $p$ is minimal. For
the case of the first model, this means that $p$ corresponds to an irreducible
representation of the symmetry group.  For the second model, we can formally regard
$p$ as an irreducible representation of some group $G$.  In category theory, we
call such a minimal $p$ as a simple object.  In this paper, we also call $p$ as
a simple particle.  

There is always a trivial simple particle denoted by $\one$. It corresponds to
local excitations that can be created by local symmetric operators
in the first model or local operators in the second model.
Its fusion space has a property
\begin{align}
\label{fuseone}
 \cV(M,\one,p,q,\cdots)\cong \cV(M,p,q,\cdots).
\end{align}

It is not hard to see that the full splitting of the fusion
space is given by (see \eqn{cVXXp})
\begin{align}
\label{cVXi}
 \cV(M,p,q,\cdots) = \cV(M,p_1,q,\cdots)\oplus \cV(M,p_2,q,\cdots) \cdots
\end{align}
In this case, we say the particle $p$ is a direct sum of particle $p_1$, $p_2$,
\etc:
\begin{align}
 p= p_1\oplus p_2 \oplus \cdots.
\end{align}
Physically, it means that the particle $p$ is an accidental degeneracy of
particle $p_1$, particle $p_2$, \etc.  For example, in the first model, we may
have a particle which is an accidental degeneracy of spin-up and spin-down
particle. Such a degeneracy becomes required in the presence of $SU(2)$ spin
rotation symmetry. In this case, a
spin-1/2 particle is a simple particle (\ie the fusion space cannot be split
further).  If we break the $SU(2)$ symmetry, then the spin-1/2 particle becomes
a composite particle which is a direct sum of two simple particles, a spin-up
and a spin-down particles.  For the case of the first model, we see that the
symmetric operations of deforming $\Del H_\text{trap}(x_p)$, which correspond
to the morphisms in category theory, allow us to define the notion of
irreducible representation without using group transformation and other
symmetry breaking operations.

\subsubsection{Fusion rules of particles}

We may view two nearby simple
particles $p_1$ and $p_2$ (\ie two irreducible representations) as one particle
$p_3$ (\ie one representation):
\begin{align}
 p_1\otimes p_2 = p_3.
\end{align}
In general $p_3$ is no longer a simple particle (\ie no longer an irreducible representation):
\begin{align}
 p_1\otimes p_2 = p_3 = p'_1\oplus p'_2\oplus \cdots.
\end{align}
Sometimes, the particle types on the right may repeat
\begin{align}
 p_1\otimes p_2=p'_1\oplus p'_1\oplus p'_2\oplus \cdots =2 p'_1\oplus p'_2\oplus \cdots .
\end{align}
We may rewrite the above as
\begin{align}
\label{fusion}
 p_i \otimes p_j = \bigoplus_k N^{ij}_k p_k,
\end{align}
which is called the fusion rule of the (simple) particles.  
From \eqn{fuseone}, we see that the trivial particle $\one$ is the
unit of the fusion operation:
\begin{align}
 \one\otimes p = p\otimes \one = p.
\end{align}
Using $N^{ij}_k$ we can calculate dimension of the fusion space
with $n$ $p_i$ particles on $S^3$, which has a form
\begin{align}
\text{dim} \cV(S^3, p_i,p_i,\cdots,p_i) =  \text{dim}\cV(S^3, p_i^{\otimes n})
\sim d_i^n
\end{align}
in the $n\to \infty$ limit. The number $d_i$ is called the quantum dimension of
the $p_i$ particle.  One can show that $d_i$ is the largest positive eigenvalue
of matrix $N_i$, where the matrix elements of $N_i$ is given by
$(N_i)_{jk}=N^{ij}_k$.

For the case of the first example, \eqn{fusion} correspond to the decomposition
of tensor product of irreducible representations.  We see that additional
information about the symmetry group $G$, the decomposition of tensor product
of irreducible representations, can also be obtained from symmetric operations:
the fusion of particles (which is realized by bring two symmetric traps
together). From $N^{ij}_k$, we can even obtain the dimensions of irreducible
representations $p_i$, which are given by the quantum dimensions $d_i$. This in
turn determines the number of elements in the symmetry group $G$:
\begin{align}
 \sum_{i \text{ is simple}} d_i^2 =|G| .
\end{align}
We get more information about the group without using any symmetry breaking
operations.

\subsubsection{Braiding and topological spin of particles}
Consider a fusion space $\cV(M,p,q,\cdots)$. If we adiabatically exchange the
two particles $p,q$, the resulting fusion space $\cV(M,q,p,\cdots)$ is always
isomorphic to the original one, no matter what the manifold $M$ and
background particles/strings are. Therefore, we say that there is a braiding
morphism $c_{p,q}$ for the fusion $p\otimes q$,
\begin{align}
  c_{p,q}: p\otimes q\cong q\otimes p.
\end{align}

In general we need to specify the exchange path (for example, clockwise or
counter-clockwise in 2+1D). But for the above two physical
models, braiding is in fact path independent. This is the defining property of
SFC, that for all particles $p,q$,
\begin{align}
  c_{q,p}c_{p,q}=\id_{p\otimes q}.
\end{align}
This means that braiding $p$ a whole loop around $q$ is the same as doing
nothing, which is equivalent to path independence.

We can also extract the \emph{topological spin} of simple particle $p$. Given a
fusion space $\cV(M,p,\cdots)$, we twist $p$ by $2\pi$, the fusion space then
acquires a phase factor $\theta_p$, called the topological spin of $p$.
It is in fact determined by the braiding $c_{p,p}$.
In the case of SFC, $\theta_p$ helps to distinguish bosons and fermions
\begin{align}
  \theta_p= \begin{cases}
   \hfill 1,& p \text{ is a boson,}\\
  -1,& p \text{ is a fermion.}
\end{cases}
\end{align}

\subsubsection{Physical realization of fiber functor}

  The Tannaka duality requires a fiber functor, which associates a
  vector space
  $F(p)$ to a particle $p$, such that it realizes the fusion and braidings of
  particles, in terms of the tensor product and the (trivial) braiding of
  vector spaces,
  \begin{align}
    F(p\otimes q)&\cong F(p)\otimes_\C F(q),\nonumber\\
    F(c_{p,q})&=c_{F(p),F(q)}.
    \label{fiberfunc}
  \end{align}
  as if $F(p)$ are local Hilbert spaces.
  Here the braiding for vector spaces is the usual one:
  \begin{align}
    c_{V,W}:v\otimes_\C w\mapsto w\otimes_\C v, \forall v\in V,w\in W.
  \end{align}
  We note that if a functor preserves the fusion (it is a monoidal functor),
  whether preserving braiding or not is just a \emph{property} of the monoidal
  functor, not an additional structure (like being an Abelian group or not is a
  property of a group). 

  We see a necessary condition for the
fiber functor to exist is that particles are all bosons with trivial braiding. It turns out
that it is also sufficient.

Physically, only the operations on the fusion spaces are measurable (or
physical).  So the question is, which fusion space should be associated to the
particle $p$ in order to have a fiber functor?  One might naturally choose the
fusion space to be $\cV(S^3,p)$ (\ie the fusion space of a particle $p$ on the
space of a 3-sphere $S^3$).  But $\cV(S^3,p)=\emptyset$ for a non-trivial
particle.  So we need to add  (non-simple) background particles and strings to
make the fusion space non-zero for any added particles.  The question is what
background particles and strings should we insert besides $p$, to get a fusion
space satisfying the conditions \eqref{fiberfunc}.

It turns out, we do have a special background (non-simple) particle to achieve
this. Let's denote it by $Q$, which has a direct sum decomposition in terms of
the simple particles and their quantum dimensions $d_i$: 
\begin{align}
    Q=\bigoplus_i d_i p_i.
\end{align}
The fusion space $\cV(S^3,p,Q)$ (no strings) satisfies
\begin{align}
    \cV(S^3,p\otimes q, Q)\cong \cV(S^3,p,Q)\otimes_\C\cV(S^3,q,
    Q).\label{fsqQ}
\end{align}
(In the first example, $Q$ is nothing but
the reducible representation
$\mathrm{Fun}(G)$, all the functions on $G$. It is the regular representation
of $G$.) Therefore, we can take
\begin{align}
F(p)\equiv \cV(S^3,p,Q).
\end{align}
It preserves fusion by \eqref{fsqQ} and also braiding (its property but we will
not show explicitly here), thus a desired fiber functor.

\subsubsection{Automorphism of the fiber functor}

Now we have a fiber functor that maps every particle $p$ to a vector space
$F(p)=\cV(S^3,p,Q)$.  Physically, the vector space $F(p)=\cV(S^3,p,Q)$ is the
ground state subspace of a Hamiltonian on $S^2$ with two traps: $H_0+\Del H_p
+\Del H_Q$, where $\Del H_Q$ traps a particular composite particle
$Q=\bigoplus_i d_i p_i$ (a particle with accidental degeneracy).

Next we like to describe the automorphism of the fiber functor.  An
automorphism is a choice of an unitary map on $F(p)=\cV(S^3,p,Q)$ for each
particle $p$.  We denote those unitary maps by $\al_p$.  So an automorphism
corresponds to a set of unitary maps $\al \equiv \{\al_p\}$.  But not every set
of unitary maps, $\{\al_p\}$, is an  automorphism.  An automorphism also needs
to preserve all the structures of the fiber functor, and as a result, needs to
satisfy many conditions.  But what are those conditions? 

We have explained that deforming the trap potential $\Del H_p$
(while preserving the symmetry in the first model) may split that fusion space
$\cV(S^3,p,Q)=\cV(S^3,p',Q)\oplus \cdots$.  This leads to a morphism
$p'\to p$.  Under the fiber functor $F$ which takes a special fusion space, the
morphism $p'\to p$ gives rise to an embedding map $u: F(p') \to F(p)$.  An
automorphism $\al=\{\al_p\}$ must be compatible with all those embedding maps:
\begin{align}
u \al_{p'}  = \al_p u,
\end{align}
or
\begin{align}
  \xymatrix{
F(p') \ar[r]^{\alpha_{p'}} \ar[d]^{u} & F(p') \ar[d]^{u}\\
  F(p) \ar[r]^{\alpha_{p}}& F(p) } .
  \label{cdrepG}
\end{align}
The map $u$ is an intertwiner.  Intertwiners are simply the
local (symmetry preserving) operations.  

In the first model, $F(p)$ is in general a reducible representation of the
symmetry group $G$.  When $p'$ is a simple particle, all the intertwiners $u$
tell us all different ways to embed irreducible representation $F(p')$ into the
reducible one $F(p)$.  The condition \eqn{cdrepG} tells us that $\al_p$ is
block diagonal and fully determined by its components on different simple
particles (irreducible representations) $\alpha_{p'}$.

The automorphism $\al=\{\al_p\}$ also needs to be compatible with the fusion of
particles.  We may view two well separated particles $p_1$ and $p_2$ as a
single particle $p_3=p_1\otimes p_2$.  The unitary maps $\al_{p_1}$,
$\al_{p_2}$, and $\al_{p_3}$ should be related.
Since the fusion space from the fiber functor satisfy
\eqn{fiberfunc},
we require  $\al_{p_3}$ equals the tensor product of
$\al_{p_1}$ and
$\al_{p_2}$ (up to the isomorphism fixed by the fiber functor
\eqn{fiberfunc}):
\begin{align}
  \xymatrix{
F(p_1\otimes p_2) \ar[r]^{\alpha_{p_1\otimes p_2}} \ar[d]^{\cong} & F(p_1\otimes p_2) \ar[d]^{\cong}\\
  F(p_1)\otimes_\C F(p_2) \ar[r]^{\alpha_{p_1} \otimes_\C \al_{p_2}} &  F(p_1)\otimes_\C F(p_2)} .
  \label{p1p2F1}
\end{align}
Since $p_3=p_1\otimes p_2=\bigoplus_i p'_i$ and $F(p_1\otimes p_2)\cong
\bigoplus_i  F(p'_i)$, the above can be rewritten as
\begin{align}
  \xymatrix{
\bigoplus_i  F(p'_i) \ar[r]^{\bigoplus_i  \al_{p'_i}} \ar[d]^{\cong} & \bigoplus_i  F(p'_i) \ar[d]^{\cong}\\
  F(p_1)\otimes_\C F(p_2) \ar[r]^{\alpha_{p_1} \otimes_\C \al_{p_2}} &  F(p_1)\otimes_\C F(p_2)} .
  \label{p1p2F}
\end{align}
The above is the condition for the automorphism $\al=\{\al_p\}$ to be compatible
with the fusion which is a data in $\Rep(G)$.  

The set of unitary maps $\al=\{\al_p\}$ that satisfies \eqn{cdrepG} and
\eqn{p1p2F} is called an automorphism of the fiber functor.  If $\al=\{\al_p\}$
and $\al'=\{\al_p'\}$ are two automorphisms, we can show that $\al\cdot
\al'\equiv\{\al_p \al_p'\}$ is also an automorphism.  So the automorphisms form
a group $G\equiv \Aut(F)$.  Such a group corresponds to the symmetry group in
the first physical model. We have measured the symmetry group using only
symmetric probes.  In the second physical model, $G$ is a group associated with
the 3+1D topological order. We have shown that every 3+1D topological order is
associated with an unique group $G$.

To emphasize the group nature of the automorphisms $\al\equiv\{\alpha_p\}$, we
may instead write $g\equiv\{g_p\}\in G\equiv \Aut(F)$. They give rise to the
group action on $F(p)$, by $\rho_p(g)=g_p$.

\subsection{Example of Tannaka reconstruction for $\Rep(Z_2)$}

In this section we illustrate the Tannaka duality with the simplest example,
$\Rep(Z_2)$.  We will follow the general reconstruction procedure, trying to
show the flavor of the abstract theorem.

Firstly let's describe $\Rep(Z_2)$ in terms of fusion. There are two
irreducible representations of $Z_2$: the trivial denoted by $\one$, the
non-trivial one denoted by $e$.  The fusion rule is
\begin{align}
  \one\otimes \one=\one,\ \ \one\otimes e=e\otimes \one=e,\ \ e \otimes e=\one.
\end{align}
The back ground charge is $Q=\one \oplus e$.  We find that $F(e)=\cV(S^3,
e\otimes Q)=\cV(S^3, e\oplus \one)=\cV(S^3, \one)=\cV(S^3)=\C$.  The ground state
on $S^3$ is non degenerate, thus $F(e)$ is one dimensional. Similarly,
$F(\one)$ is  one dimensional as well.

When $p$ is composite, $p=\bigoplus_i p_i$, \eqn{cdrepG} tells us that $\al_p$
is block diagonal
\begin{align}
 \al_p = \bigoplus_i\al_{p_i},
\end{align}
where $p_i$ are simple.  Since $F(p_i)$ for a simple particle is always
one dimensional for $\Rep(Z_2)$, $\al_\one$ and $\al_e$ are just phase factors.
Eqn. (\ref{p1p2F}) requires that
\begin{align}
\al_{\one\otimes e} =  \al_\one\otimes_\C \al_e=\al_e.
\end{align}
Thus $\al_\one=1$.
Eqn. (\ref{p1p2F}) also requires that
\begin{align}
\al_{e\otimes e} = \al_e\otimes_\C \al_e= \al_\one=1.
\end{align}
Thus $\al_e=\pm 1$.  The solution $\{\al_\one=1,\al_e=1\}$ corresponds to an
automorphism, and the solution $\{\al_\one=1,\al_e=-1\}$ corresponds to the
other automorphism.  The composition $\{\al_\one,\al_e\}\{
\al_\one',\al_e'\}=\{\al_\one\al_\one',\al_e\al_e'\}$ is the group
multiplication, which tells us that $\{\al_\one=1,\al_e=1\}$ and
$\{\al_\one=1,\al_e=-1\}$ form a $Z_2$ group.
\subsection{Tannaka duality II: with fermions}
We proceed to introduce the Tannaka duality for SFC $\sE$ which contains
fermions. The idea is almost the same: find a fiber functor, calculate the
automorphisms of the fiber functor, and we recover the group. But the fiber
functor needs to preserve braiding, while in $\Ve$ there are only bosons. So
we have to change the target of the fiber functor to accommodate fermions. The
new target category is just the simplest SFC that contains fermions,
namely the category of super vector spaces $\sVe$.
The fusion part of $\sVe$ is the same as
$\Rep(Z_2)$. But now the non-trival particle, denoted by $f$ to distinguish
from the $\Rep(Z_2)$, is a fermion;
its braiding is modified:
\begin{align}
  c_{f,f}=-\id_{\one}.
\end{align}
while other braidings remain trivial. It can be understood as vector spaces
with a $Z_2$ grading. The non-trivial grading corresponds to fermionic degrees
of freedom, while the trivial grading corresponds to bosonic degrees of freedom.

So when there are fermions in $\sE$, we instead need a super fiber functor
\begin{align}
  F:\sE\to \sVe,
\end{align}
It can be physically realized the same way using the fusion space $\cV(S^3,q,Q)$.
And we can follow exactly the same procedure introduced in the last subsection
to construct a group from automorphisms of the super fiber functor $F$,
\begin{align}
  G_f\equiv \Aut(F).
\end{align}
Such a group is slightly different from the bosonic case.
Note that there is a special automorphism $z=\{z_p\}$,
\begin{align}
  z_p=
  \begin{cases}
    \hfill \id_{F(p)},& p\text{ is a boson,}\\
    -\id_{F(p)},& p\text{ is a fermion.}
  \end{cases}
  \label{parity}
\end{align}
$z$ corresponds to the fermion number parity and commutes with all other
automorphisms. Let $Z_2^f\equiv\{1,z\}$. We see that the group $G_f$ must contain
$Z_2^f$ as a central subgroup. We then have
\begin{align}
  \sE\cong \sRep(G_f).
\end{align}
Where $\sRep(G_f)$ is constructed similarly like $\Rep(G)$. They have the same
fusion; only the braiding between two fermions has an extra $-1$. In this
sense we have $\sVe=\sRep(Z_2^f)$. 

\subsection{(Super) fiber functor from condensation}
In the above we realized the (super) fiber functor using the fusion space on
$S^3$ with
a special background particle $Q$. But we gave no proof why such fusion space
preserves the fusion and braiding. In this subsection we give a physical reason why such $Q$
is so special.

In the all-boson case, imagine that we let $Q$ condense to form a new phase, a \emph{$Q$-sea}, such
that $Q$ becomes the trivial particle in the $Q$-sea. One expects
the fusion space to remain the same,
\begin{align}
  \cV(S^3,p,Q)&=\cV(S^3,p,\text{trivial particle above }Q\text{-sea})\nonumber\\
  &=\cV(S^3,p,Q\text{-sea}).
\end{align}
So the properties of $\cV(S^3,p,Q)$ in fact follows from those of the $Q$-sea,
as in $\cV(S^3,p,Q\text{-sea})$, the particle $p$ behaves like a particle above
the $Q$-sea.
Then it is clear that we want the $Q$-sea to be a trivial phase, whose
particles are described by $\Ve$.

If there are fermions, similarly we want a condensate whose particles form
$\sVe$. But $Q$ should become, instead of the trivial particle, a direct sum
$\one\oplus f$, from whose fusion space we can extract both bosonic and
fermionic degrees of freedom. It turns out $Q$ should be of the following form:
\begin{align}
  Q=Q_b\oplus Q_f,\quad \dim(Q_b)=\dim(Q_f),
\end{align}
where $Q_b$ and $Q_f$ are bosonic and fermionic parts respectively. We condense
the bosonic part $Q_b$, and particles above the $Q_b$-sea should be $\sVe$,
\begin{align}
  \cV(S^3,p,Q)&=\cV(S^3,p,\one\oplus f\text{ above }Q_b\text{-sea}).
\end{align}

It is indeed from these requirements on the condensation how we determine the special
particle $Q$. This idea of condensation is also the main physical motivation of
this paper.

\section{Relation between emergent Majorana zero modes for linked loops
and the 2-cocycle $\rho_2$}

\begin{figure}[t]
  \centering
  \includegraphics{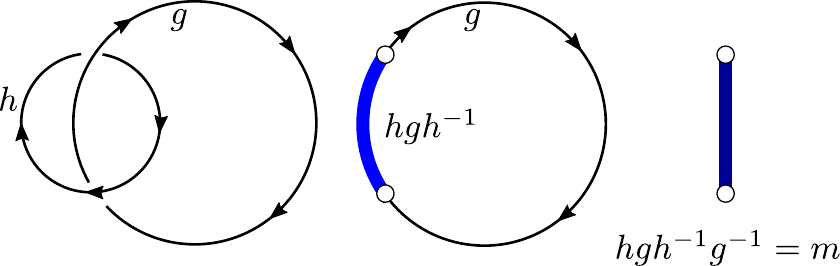}
  \caption{Fuse $h$ loop to the linked $g$ loop on the canonical domain wall.
    When $hgh^{-1}=gm$ two Majorana zero modes are supported. Further fusing
  the two segments we obtain an open Majorana chain.}
  \label{linkM}
\end{figure}

In \Ref{CJWang}, it was pointed out that, for some fermionic SPT states,
certain linked loops of symmetry twists can carry a pair of Majorana zero modes
(see Fig. \ref{linkM}).  In this section, we like to discuss a relation between
such emergent Majorana zero modes and the non-trivial two cocycle $\rho_2$ that
characterize the EF2 topological orders.  For simplicity, we assume $G_f$ to be
Abelian.  We will show that certain linked looplike excitations in a EF2
topological order carry a pair of Majorana zero modes, one for each linked
loop. In other words, certain pairs of looplike excitations carry two-fold
topological degeneracy when they are linked and no degeneracy when they are not
linked.  Such a topological degeneracy is highly non-local in the sense that
the degeneracy is shared between the two well separated linked loops.  The new
result here is that the appearance of  Majorana zero modes for linked loops is
directly related to the non-trivial $Z_2^m$ extension of $G_b$ on the canonical
boundary.

To see the above result, we consider a pair of linked loops in the bulk in Fig.
\ref{linkM}.  We know that a pair of linked loops in the bulk is characterized
by a pair of commuting elements $h^f,g^f$ in $G_f$ (assuming $G_f$ is
non-Abelian for the moment).  (To be more precise, a pair of linked loops is
characterized by the conjugacy class of a pair of commuting elements $h^f,g^f$.)
As we go around a loop, the string labeled by $g^f$ is changed into the  string
labeled by $h^fg^f(h^f)^{-1}$. The string can form a loop only when
$g^f=h^fg^f(h^f)^{-1}$. It is why $h^f,g^f$ describing linked loops must
commute.  

Now, let us assume $G_f$ is Abelain.  We like to compute the degeneracy for the
linked loops in Fig.  \ref{linkM}.  For Abelian $G_f$, all the
pointlike excitations and stringlike excitations have an unit quantum
dimension.  Thus one may expect that  degeneracy for the linked loops to be 1.
In the following, we like to show that some times the degeneracy can be 2.  To
obtain such a result, we bring the linked loops to the boundary.  This reduces
the group elements $h^f,g^f$ in $G_f$ to the group elements $h^b=\pi^f(h^f),\
g^b=\pi^f(g^f)$ in $G_b$ via the natural reduction $G_f \stackrel{\pi^f}{\to}
G_b=G_f/Z_2^f$.  In addition to the reduction $G_f \to G_b$, there is also an
extension $G_b \to \hat G_b$.  So the linked loops on the boundary are actually
described by $h,g$ in $\hat G_b$, where $ h^b=\pi^m(h),\ g^b=\pi^m(g) $ under
the projection $\hat G_b \stackrel{\pi^m}{\to} G_b$.  To summarize, the bulk
string types $h^f,g^f$ turn to boundary string l-types $h,g$ that satisfy the
following relation
\begin{align}
 \pi^f(g^f) = \pi^m(g), \ \ \ \ \ \pi^f(h^f) = \pi^m(h).
\end{align}
This is the situation described in Fig. \ref{linkM}.  As we go around a loop,
boundary string labeled by $g$ turns into a boundary string $h g h^{-1}$.  Even
though $h^b,g^b$ commute in $G_b$, their lifts $h,g$ may not commute in
$\hat G_b$, when $\hat G_b$ is a non-trivial $Z_2^m$ extension of $G_b$.  If
$h,g$ do not commute, we will have $h g h^{-1}=gm$ where $m$ generates $Z_2^m$.
As a result, there are two  pointlike defects between $g$ and $gm$
boundary strings, corresponding to two Majorana zero modes which lead to a
2-fold degeneracy.

To see which linked loops described by $h^f,g^f$ have Majorana zero modes, we
first note that the elements in $\hat G_b$ can be labeled as $(g^b,x)$, $g^b
\in G_b$ and $x \in Z_2^m$.  The multiplication in $\hat G_b$ is given by
\begin{align}
 (g^b,x) (h^b,y)=(g^bh^b, \rho_2(g^b,h^b) xy)
\end{align}
where $\rho_2(g^b,h^b)$ is the group 2-cocycle in $H^2(G_b,Z_2^m)$.  For
$h^f,g^f$, we have $h=(\pi^f(h^f),y), g=(\pi^f(g^f),x) \in \hat G_b$. As shown in
Fig.~\ref{linkM}, their
commutator $[h,g]\equiv hgh^{-1}g^{-1}=hg(gh)^{-1}$ determines the appearance of Majorana
zero modes. Without losing generality, we may assume that $\rho_2$ is a
normalized 2-cocycle, namely $\rho_2(1,g^b)=\rho_2(g^b,1)=1$, $\forall g^b\in
G_b$. Using the fact that $hg=[h,g]gh$ and
$\pi^f(h^f)\pi^f(g^f)=\pi^f(g^f)\pi^f(h^f)$, it is easy to compute
$[h,g]=(1,\rho_2(\pi^f(h^f),\pi^f(g^f))\rho_2(\pi^f(g^f),\pi^f(h^f)))$.  We see that the
linked loops $h^f,g^f$ have Majorana zero modes when
$\rho_2(\pi^f(h^f),\pi^f(g^f))\rho_2(\pi^f(g^f),\pi^f(h^f))=m$.  The
appearance of Majorana zero modes for certain linked loops can detect a certain
type of non-trivial $Z_2^m$ extensions, \ie those with non-trivial
$\rho_2(\pi^f(h^f),\pi^f(g^f))\rho_2(\pi^f(g^f),\pi^f(h^f))$ for certain pairs of
elements $h^f,g^f$ in $G_f$.

\bibliography{local,./wencross,./all,./publst}

\end{document}